\documentclass{ws-ijmpd}
\usepackage[super,compress]{cite}
\begin{document}

\markboth{Herdeiro and Radu}
{Asymptotically flat black holes with scalar hair: a review}

\catchline{}{}{}{}{}
%
%%%%%%%%%%%%%%%%%%%%%%%%%%%%%%%%%%%%%%%%%%%%%%%%%%%%%%%%%%%%%%%%%%%%

\title{Asymptotically flat black holes with scalar hair: \\ a review}

\author{Carlos A. R. Herdeiro and Eugen Radu}

\address{Departamento de F\'\i sica da Universidade de Aveiro and CIDMA, \\ Campus de Santiago, 3810-183, Aveiro, Portugal\\ 
herdeiro@ua.pt, eugen.radu@ua.pt}

\maketitle

\begin{history}
%\received{Day Month Year}
%\revised{Day Month Year}
\end{history}

\begin{abstract}
We consider the status of black hole solutions with non-trivial scalar fields but no gauge fields, in four dimensional asymptotically flat space-times, reviewing both classical 
results and recent developments. We start by providing a simple illustration 
on the physical difference between black holes in electro-vacuum and scalar-vacuum. 
Next, we review no-scalar-hair theorems. In particular, we detail an influential theorem by Bekenstein and stress three key assumptions: 1) the type of scalar field equation; 2) the spacetime symmetry inheritance by the scalar field; 3) an energy condition. Then, we list regular (on and outside the horizon), asymptotically flat BH solutions with scalar hair, organizing them by the assumption which is violated in each case and distinguishing primary from secondary hair. We provide a table summary of the state of the art.
\end{abstract}

\keywords{black holes; scalar fields; no-hair theorems}

\ccode{PACS numbers: 04.70.Bw; 04.20.Jb}

\newpage

\tableofcontents

\section{Introduction}	
In (electro-)vacuum general relativity, equilibrium black holes (BHs) are very special objects. Other celestial bodies, say two stars, with the same total mass $M$ and angular momentum $J$ can be very different, since these quantities can be differently distributed within the star; but two BHs under these circumstances will be \textit{exactly} equal. Indeed, the most general regular solution, on and outside an event horizon, is the Kerr(-Newman) metric, as established by the \textit{uniqueness theorems}~\cite{Robinsoon:2004zz,Chrusciel:2012jk}, for which all higher multipole moments are determined by only $M,J$ (and electric charge, if present).\footnote{This statement applies to single-BH solutions. Multi-BH solutions, described by the Majumdar-Papapetrou metric, also exist in electro-vacuum. Moreover, if one is willing to consider magnetic charges, by electromagnetic duality there are also magnetically charged and dyonic BHs. None of these, however, seem likely to arise as the endpoint of gravitational collapse.} These uniqueness theorems are sometimes referred to as \textit{no-hair theorems}, a naming that should be understood as ``no-independent-multipole-hair": higher gravitational multipole moments -- quadrupole and higher -- (and electromagnetic multipole moments -- dipole and higher) are not independent for electro-vacuum BHs.

The uniqueness theorems led to the conjecture that the outcome of gravitational collapse in the presence of \textit{any type} of matter-energy is a Kerr-Newman BH, solely described by mass, angular momentum and electric charge, all of these asymptotically measured quantities \textit{subject to a Gauss law}, and no other physical quantities, to which ``hair" provides a metaphor, should exist. This is the \textit{no-hair conjecture}.\cite{Ruffini:1971bza} Observe that here, ``hair" encodes a much broader use than that in the previous paragraph.

As just stated, the no-hair conjecture concerns the \textit{dynamical} end point of gravitational collapse, and not just the mere existence of a stationary BH solution with some type of matter-energy, regardless of its dynamics and, in particular, of its stability. Throughout the years, however, stationary BH solutions with either new global charges (\textit{primary} hair) or new non-trivial fields -- even if not independent from the standard global charges (\textit{secondary} hair) -- which are not associated to a Gauss law, have been generically referred to as `hairy BHs'. In the 1980s and 90s a variety of hairy BH solutions have been found, typically in theories with non-linear matter sources.\cite{Bizon:1994dh,Bekenstein:1996pn,Volkov:1998cc} But, in the late 1990s, this was not the case for scalar field hair. Indeed, as observed by Mayo and Bekenstein:\cite{Mayo:1996mv} {\it
``The proliferation in the 1990Õs of stationary black hole solutions with  hair  
of various sorts may give the impression that the principle has fallen by the wayside. 
However, this is emphatically not the case for scalar field hair." 
}

Scalar fields are one of the simplest types of ``matter'' often considered by physicists. 
Moreover, since 2012, there is observational evidence that fundamental scalar fields exist in nature, by virtue of the discovery of a scalar particle at the Large Hadron Collider, at CERN, identified as the standard model Higgs boson~\cite{Aad:2012tfa,Chatrchyan:2012ufa}.
But for decades, scalar fields have been considered in phenomenological models, in particular within gravitational physics.
A notable example is cosmology, where various types of scalar fields have been used to model dark energy and dark matter.
One reason is that scalar fields are well motivated by \textit{beyond the standard model particle physics}, both as fundamental fields and as effective fields arising as a coarse-grained description of more fundamental fields. Yet another reason is that scalar fields may be considered as a proxy to realistic matter, since canonical scalar fields can be modeled as perfect fluids with some equation of state.\cite{Faraoni:2012hn}

Why should scalar fields be different from electromagnetic fields, in terms of BH hair? A simple illustration is provided in Section~\ref{sec_2}, by comparing electro-vacuum with scalar-vacuum general relativity, and an essential difference is the existence of a Gauss law for the former but not for the latter. In any case, due to their simplicity,  it is quite natural that in testing the no-hair idea, scalar fields were one of the first types of ``matter'' considered. A set of no-go results -- \textit{no-scalar-hair theorems} -- will be reviewed in Section~\ref{sec_3}, where we shall emphasize the assumptions that go into an influential example of a no-hair theorem due to Bekenstein\cite{Bekenstein:1972ny}. But we shall also review other no-scalar-hair theorems and, in particular, discuss variations of  the energy assumption that goes into the simplest Bekenstein-type theorem. Then, in Section~\ref{sec_4} we shall review solutions of asymptotically flat BHs with scalar hair, organizing them in terms of the assumption of no-scalar-hair theorems they violate. In particular we shall review some recently found solutions with qualitatively different properties. The state of the art is summarized  in Section~\ref{sec_5}. 
Finally, in Section~\ref{sec_6} we provide some final remarks.

%%%%%%%%%%%%%%%%%%%
\section{Electro-vacuum $vs.$ scalar-vacuum: a simple illustration}
\label{sec_2}
%%%%%%%%%%%%%%%%%%%
Some simple observations show the distinction between trying to superimpose an electric field and a scalar field on a BH spacetime. For this purpose we contrast electro-vacuum, described by the action ($G=c=4\pi\epsilon_0=1$)
\begin{equation}
\mathcal{S}=\frac{1}{4\pi }\int d^4x\sqrt{-g}\left(\frac{R}{4}-\frac{1}{4}F_{\mu\nu}F^{\mu\nu}\right) \ ,
\label{ev_action}
\end{equation}
and scalar-vacuum, for a massless, real scalar field, described by the action
\begin{equation}
\mathcal{S}=\frac{1}{4\pi }\int d^4x\sqrt{-g}\left(\frac{R}{4}-\frac{1}{2}\nabla_{\mu}\Phi \nabla^\mu\Phi\right) \ .
\label{sv_action}
\end{equation}
In both cases, the Schwarzschild metric of vacuum general relativity, with mass $M$, is a solution (with $F_{\mu\nu}=0$ and $\nabla_\mu\Phi=0$, respectively). In standard Schwarzschild coordinates it reads:
\begin{equation}
ds^2=-\left(1-\frac{2M}{r}\right)dt^2+\frac{dr^2}{1-2M/r}+r^2(d\theta^2+\sin^2\theta d\phi^2) \ .
\label{sch}
\end{equation}
Linearizing the field equations of \eqref{ev_action} (or \eqref{sv_action}) in $F$ (or $\Phi$) around the Schwarzschild metric, yields the source free Maxwell (or Klein-Gordon) eq. in the background \eqref{sch}.

%%%%%%%%%%%%%%%%%%%
\subsection{Spherically symmetric fields}
\label{sec_21}
%%%%%%%%%%%%%%%%%%%
As a first example that illustrates the difference between the two types of fields, consider on the Schwarzschild background \eqref{sch} a test, spherically symmetric: $i)$ electric field, described by the gauge potential $A=\phi_E(r)dt$ and corresponding Maxwell tensor $F=dA$; $ii)$ scalar field, described by the radial profile $\Phi(r)$. From the source free Maxwell and Klein-Gordon equations one obtains:
\begin{equation}
D_\mu F^{\mu \nu}=0 \ \Rightarrow \ \partial_r\phi_E(r)=\frac{Q_E}{r^2} \ \Rightarrow \ \phi_E(r)=-\frac{Q_E}{r} \ , 
\label{pot_l}
\end{equation}
\begin{equation}
\Box \Phi(r)=0 \ \Rightarrow \ \partial_r\Phi(r)=\frac{Q_S}{r^2}\left(1-\frac{2M}{r}\right)^{-1} \ \Rightarrow \ \Phi(r)=\frac{Q_S}{2M}\ln\left(\frac{2M}{r}-1\right) \ , 
\label{testscalar}
\end{equation}
where $Q_E,Q_S$ are integration constants. In the first case one obtains a solution which is \textit{regular} on and outside the horizon. This electric field, moreover, sources a regular energy--momentum on and outside the horizon, 
\begin{equation}
T_{\mu\nu}^E=
F_{\mu\alpha}F_{\nu}^{\ \alpha}-\frac{1}{4}g_{\mu\nu}F_{\alpha\beta}F^{\alpha\beta},
\end{equation}
with non-trivial components:
\begin{equation}
(T^E)^{t}_{\ t}=(T^E)^r_{\ r}=-\frac{Q_E^2}{2 r^4}=-(T^E)^{\theta}_{\ \theta}=-(T^E)^\phi_{\ \phi} \ .
\end{equation}
 By making this electric field backreact on the metric one obtains, solving the electro-vacuum Einstein equations $G_{\mu\nu}= 2 T_{\mu\nu}^E$, the Reissner-Nordstr\"om BH solution:
 \begin{equation}
ds^2=-\left(1-\frac{2M}{r}+\frac{Q_E^2}{r^2}\right)dt^2+\frac{dr^2}{1-2M/r+Q_E^2/r^2}+r^2(d\theta^2+\sin^2\theta d\phi^2) \ ,
\label{RN}
\end{equation}
\begin{equation}
A=-\frac{Q_E}{r}dt \ .
\label{pot_nl}
\end{equation}
Curiously, for the non-linear solution, the form of the electromagnetic potential \eqref{pot_nl} is the same as for the test field ($i.e.$ linear) solution \eqref{pot_l}. The RN solution continuously connects with the Schwarzschild solution and has a regular horizon as long as the constant $Q_E$ obeys $|Q_E|<M$. Moreover, this constant can be computed as the electric flux  on a closed 2--surface $\partial \Sigma$, with area element $dS_{\mu\nu}$:
\begin{equation}
Q_E=\frac{1}{8\pi}\oint_{\partial\Sigma} F^{\mu\nu}dS_{\mu\nu} \ ;
\label{gauss}
\end{equation}
due to the spherical symmetry we choose $\partial \Sigma$ as $r,t$=constant surface at any $r$ outside the BH. Thus $Q_E$ is the electric charge, which obeys a Gauss law.

In the second case, the scalar field gradient \textit{diverges} as $(r-r_H)^{-1}$ at the horizon $r_H=2M$. Thus the scalar field diverges logarithmically therein. More importantly, the scalar energy--momentum tensor 
 \begin{equation}
T_{\mu\nu}^S=\partial_\mu\Phi\partial_\nu\Phi-\frac{1}{2}g_{\mu\nu}\partial_\alpha\Phi\partial^\alpha\Phi \ ,
 \end{equation}
 also diverges on the horizon:
 \begin{equation}
(T^S)^r_{\ r}=\frac{Q_S^2}{2r^4}\left(1-\frac{r_H}{r}\right)^{-1}=- (T^S)^{t}_{\ t}=-(T^S)^{\theta}_{\ \theta}=-(T^S)^\phi_{\ \phi} \ .
 \end{equation}
 Firstly, this shows that the test field approximation for the scalar field always fails near the horizon, no matter how small $Q_S$ is. Secondly, this provides evidence, that no regular (on and outside a horizon), spherically symmetric and static solution of a BH with scalar hair exists, connecting continuously to the Schwarzschild solution. But actually, a solution to the scalar-vacuum Einstein equations $G_{\mu\nu}=2 T_{\mu\nu}^S$ \textit{does exist}, which reduces to the above test field analysis in the limit of small enough scalar field in most of the spacetime. The solution was found by Fisher~\cite{Fisher:1948yn} and independently rediscovered by Janis, Newman and Winicour~\cite{Janis:1968zz} and reads:
\begin{equation}
ds^2=-\left[\frac{R-M(\mu-1)}{R+M(\mu+1)}\right]^{1/\mu}dt^2+\left[\frac{R+M(\mu+1)}{R-M(\mu-1)}\right]^{1/\mu}dR^2+r(R)^2(d\theta^2+\sin^2\theta d\phi^2) \ ,
\label{jnw}
\end{equation}
\begin{equation}
\Phi(R)=\frac{Q_S}{2M\mu}\ln\left[\frac{R-M(\mu-1)}{R+M(\mu+1)}\right]\ ,
\label{sca_nl}
\end{equation}
where the areal radius $r$ is a function of the radial coordinate $R$:
\begin{equation}
r(R)^2=\left[{R-M(\mu-1)}\right]^{1-1/\mu}\left[R+M(\mu+1)\right]^{1+1/\mu} \ ,
\label{arealradius} 
\end{equation}
and the parameter $\mu$ measures the non-linearity of the scalar field:
\begin{equation}
\mu\equiv \sqrt{1+ \frac{Q_S^2}{M^2}} \ .
\end{equation}
In particular if one takes $Q_S/M\ll 1$, so that $\mu\simeq 1$ and (when $R>0$) $r\simeq R+2M$, then \eqref{jnw} becomes the Schwarzschild metric \eqref{sch} and the scalar field reduces to the test scalar field \eqref{testscalar}. This limit, however, is subtle when reaching the point where $R$=$const.$ surfaces stop being timelike. Taking $\mu\simeq 1$ this occurs for $R=0$, in which case analysing the limit of \eqref{arealradius}, shows that $r=0$, no matter how small $Q_S/M$ is. Thus, the $t,R$=$const.$ surfaces suddenly collapses from a sphere with radius slightly greater than $r=2M$ to zero~\cite{Janis:1968zz}. This is how the fully non-linear solution copes with the aforementioned observation that the test field analysis always breaks down at the horizon. In any case, the main physical feature of this solution is that the areal radius vanishes at the (would be) horizon, where the geometry possesses a physical curvature singularity.

Physically, one interpretation for the difference between the scalar and the electric cases described above, 
is related to the existence of a Gauss law for the electric field~\eqref{gauss}, which has no equivalent for the scalar field. Thus, the regular electric field on and outside the horizon can be sourced by charges that have fallen into the BH. On the other hand, for model (\ref{sv_action}), a non-trivial scalar field outside the horizon -- no matter how small at some distance from the horizon -- implies an infinite pile up on the horizon, as any finite amount of scalar field placed outside the BH as initial data should either disperse to infinity or fall into the horizon. In the latter case no trace of the scalar field remains outside the BH, due to the absence of a Gauss law. The exact non-linear solution for the scalar-vacuum system (\ref{sv_action}), 
then reveals that this infinite scalar pile up at the horizon compresses the horizon to vanishing area and renders the geometry singular therein.

As a final observation, to which we shall come back in Section~\ref{horgal}, the divergence for the \textit{test} scalar field solution, eq.~\eqref{testscalar}, on the Schwarzschild horizon can be cured by adding to it a linearly time dependent term, which also solves the Klein-Gordon equation. Thus, 
\begin{equation}
\Phi(t,r)=\frac{Q_S}{2M}\left[\frac{t}{2M}+\ln\left(\frac{2M}{r}-1\right)\right] \ .
\label{jacobson}
\end{equation}
solves $\Box \Phi=0$ on the Schwarzschild metric~\eqref{sch} and is regular on the horizon. This solution was suggested by Jacobson\cite{Jacobson:1999vr} in the context of scalar-tensor theories ($cf.$ Section~\ref{sec_drop1}). In scalar-vacuum Einstein's gravity, however, this solution is not compatible with either stationarity or asymptotic flatness.

%%%%%%%%%%%%%%%%%%%
\subsection{Non-spherically symmetric fields}
\label{nss}
%%%%%%%%%%%%%%%%%%%

Is spherical symmetry essentially for the above results? In other words are there regular (on and outside an event horizon), asymptotically flat, static BH solutions in electro--vacuum or scalar--vacuum with higher multipoles? Again, a test field analysis is informative. Allow now the above fields $\phi_E$ and $\Phi$ to have an arbitrary angular dependence obtained as a superposition of spherical harmonics $Y_{\ell}^m(\theta, \phi)$. Since the source free Maxwell and Klein-Gordon equations are linear, one may consider each harmonic term separately and solve each of these equations on the Schwarzschild background with the ansatz
\begin{equation}
\phi_E(r,\theta,\phi)=R^E_{\ell}(r)Y_{\ell}^m(\theta,\phi)  \ , \qquad \Phi(r,\theta,\phi)=R^S_{\ell}(r)Y_{\ell}^m(\theta,\phi) \ ;
\end{equation}
the equations yield:
\begin{equation}
\left(1-\frac{2M}{r}\right)\frac{d}{dr}\left[r^2\frac{dR^E_\ell}{dr}\right]=\ell(\ell+1)R^E_\ell \ ,
\end{equation}
and
\begin{equation}
\frac{d}{dr}\left[r^2\left(1-\frac{2M}{r}\right)\frac{dR^S_\ell}{dr}\right]=\ell(\ell+1)R^S_\ell \ .
\end{equation}
One can check that, for both cases, the generic solution for any $\ell\neq 0$ is a linear combination of one solution that diverges at the horizon and another solution that diverges at infinity. For instance, for $\ell=1$
\begin{equation}
R_1^E(r)=c_1(r-2M)+c_2\left[\frac{M}{r}-1+\left(1-\frac{r}{2M}\right)\ln\left(1-\frac{2M}{r}\right)\right] \ ,
\end{equation}
and
\begin{equation}
R_1^S(r)=c_1(r-M)+c_2\left[-1+\left(\frac{1}{2}-\frac{r}{2M}\right)\ln\left(1-\frac{2M}{r}\right)\right] \ ,
\end{equation}
where $c_1,c_2$ are constants. 

This analysis suggests that no static, regular (on and outside a horizon) non-spherically symmetric, asymptotically flat BH solutions exist for either electro-vacuum or scalar-vacuum, connecting continuously to Schwarzschild.\footnote{Irregular solutions can, of course, be found, using, for instance, the Weyl formalism.} In the electro-vacuum case -- and also for the scalar-vacuum case --, this may be understood along the lines of the discussion in Section~\ref{sec_21}: there is no conservation law for higher electric multipoles (in particular there is no Gauss law), unlike for the monopole (total charge). For the electro-vacuum case the inexistence of regular solutions with multipoles is proven by one of Israel's uniqueness theorems~\cite{Israel:1967za}. For scalar-vacuum, this is proven by a no-hair theorem, as we discuss in the Section~\ref{sec_3}.

 %%%%%%%%%%%%%%%%%%%
\subsection{Beyond scalar-vacuum: conformal scalar-vacuum}
\label{sec_23}
%%%%%%%%%%%%%%%%%%%
There are two important words of caution concerning the lessons of the previous subsections. Firstly, we centered the discussion around possible BH solutions with scalar hair continuously connecting to the Schwarzschild solution. Could there be scalar-hairy BHs that do not reduce to the Schwarzschild BH? Secondly, the divergence of the scalar field at the horizon translated into a divergence of the scalar energy-momentum tensor at the horizon and thus of the curvature. Could a diverging scalar field at the horizon not translate into a curvature singularity at the (would be) horizon?

The answer to both this questions is yes as it is illustrated by the much debated Bocharova--Bronnikov--Melnikov--Bekenstein (BBMB) BH solution\cite{BBM,Bekenstein:1974sf,Bekenstein:1975ts}
 of \textit{conformal scalar-vacuum}, which has the action:
 \begin{equation}
\mathcal{S}=\frac{1}{4\pi}\int d^4x\sqrt{-g}\left(\frac{R}{4}-\frac{1}{2}\nabla_{\mu}\Phi \nabla^\mu\Phi-\frac{1}{12}R\Phi^2\right) \ .
\label{csv_action}
\end{equation}
The scalar field equation, $\nabla_\mu \nabla^\mu\Phi-\Phi R/6=0$ is invariant under a local conformal transformation, $g_{\mu\nu}\rightarrow \hat{g}_{\mu\nu}=\Omega^2 g_{\mu\nu}$ and $\Phi\rightarrow \hat{\Phi}=\Phi/\Omega$ (albeit the Einstein-Hilbert term, and hence the full action, is not), which justifies the name conformal scalar-vacuum for this theory. It is a special case of scalar-tensor theories, $cf.$ Section~\ref{sec_drop1}. The BBMB solution of this theory reads:
 \begin{equation}
ds^2=-\left(1-\frac{M}{r}\right)^2 dt^2+\frac{dr^2}{\left(1-M/r\right)^2}+r^2(d\theta^2+\sin^2\theta d\phi^2) \ ,
\label{BBMB1}
\end{equation}
\begin{equation}
\Phi=\frac{\sqrt{3}M}{r-M} \ .
\label{BBMB2}
\end{equation}
This is a one-parameter family of solutions (parameter is $M$, the total mass). When $M=0$ it reduces to Minkowski space. For any other value of $M$ the geometry coincides with that of an extremal Reissner-Nordstr\"om BH, $i.e.$ eq. \eqref{RN} with $|Q_E|=M$. In particular it has a regular horizon and hence it is a BH. Thus this BH with scalar hair does not connect to Schwarzschild. Moreover, the scalar field diverges at the horizon, even though the geometry is regular therein. 

This solution has been shown to be unstable against linear perturbations\cite{Bronnikov:1978mx} (other authors, however, have made a different claim\cite{McFadden:2004ni}). This proof actually covers a slightly more general family of solutions obtained by adding a Maxwell field. As such one may argue it does not violate the dynamical spirit of the no-hair conjecture. Also, observe that there is no independent scalar charge. This type of (dependent on other global charges)  scalar hair is sometimes called \textit{secondary}, as opposed to \textit{primary} hair, which has an independent charge. This distinction is important in the context of understanding the number of independent parameters that fully characterize a BH. But even secondary hair has physical consequences if it induces a BH geometry different from those of the paradigmatic BHs of general relativity. Yet another objection to BBMB solution has been raised in connection to the properties of the energy-momentum tensor at the horizon.\cite{Sudarsky:1997te}

%%%%%%%%%%%%%%%%%%%
\section{No-scalar-hair theorems}
\label{sec_3}
%%%%%%%%%%%%%%%%%%%
Chase~\cite{Chase:1970}, following an earlier suggestion~\cite{Penney:1968zz}, 
first considered if a regular BH solution could exist in scalar-vacuum by dropping the condition of spherical symmetry. He established that ``every zero-mass scalar field which is gravitationally coupled, static and asymptotically flat, becomes singular at a simply-connected event horizon''. In other words, a static BH spacetime cannot support a regular massless scalar field in equilibrium with it; $i.e.$ no \textit{static BH (massless) scalar hair} in scalar-vacuum. Bekenstein then developed a different method for proving the inexistence of scalar hair~\cite{Bekenstein:1971hc,Bekenstein:1972ky,Bekenstein:1972ny}, which became influential, and applied it for massive scalar as well as higher spin fields. We shall now review (a slight generalization of\cite{Bekenstein:1995un}) Bekenstein's original proof\cite{Bekenstein:1972ny} to emphasize three underlying assumptions.\footnote{Independently, this same proof was also sketched in the paper of Hawking~\cite{Hawking:1972qk} on BHs in Brans-Dicke theory, $cf.$ Section~\ref{sec_drop1}.} This will help understanding how the solutions to be described in the next Section are compatible with no-hair theorems. 

%%%%%%%%%%%%%%%%%%%%
\subsection{Bekenstein's theorem for $V$-scalar-vacuum}
\label{sec_bek_th}
%%%%%%%%%%%%%%%%%%%%

Consider a rotating, stationary, asymptotically flat BH spacetime. Hawking established that, assuming the null energy condition (which is implied by the weak energy condition), the spacetime is also axi-symmetric -- \textit{rigidity theorem} -- and that the spatial sections of the horizon are topologically spheres~\cite{Hawking:1971vc}. We write the spacetime metric in coordinates adapted to these symmetries $(t,r,\theta,\phi)$, so that the two Killing vector fields read ${\bf k}=\partial_t$, ${\bf m}=\partial_\phi$.

\bigskip

{\bf Assumption 1: consider a canonical and minimally coupled scalar field to Einstein's gravity.} Allowing the possibility of a potential, $V(\Phi)$, the action is a slight generalization of \eqref{sv_action} (hereafter, this theory is dubbed \textit{V-scalar-vacuum}):
\begin{equation}
\mathcal{S}=\frac{1}{4\pi}\int d^4x\sqrt{-g}\left(\frac{R}{4}-\frac{1}{2}\nabla_{\mu}\Phi \nabla^\mu\Phi-V(\Phi)\right) \ .
\label{sv_action2}
\end{equation}
Thus the scalar field obeys the (possibly non-linear) Klein-Gordon equation:
\begin{equation}
\label{n-KG}
\nabla_\mu \nabla^\mu \Phi- V'(\Phi)=0 \ , 
\end{equation}
where the prime denotes derivative with respect to the argument. In particular this means the scalar field is minimally coupled to the geometry, and excludes from this theorem non-minimally coupled scalars, and thus scalar-tensor theories of gravity. For a non-self-interacting massive scalar with mass $\mu$, $V(\Phi)=\frac{1}{2}\mu^2\Phi^2$.

\bigskip

{\bf Assumption 2: the scalar field inherits the spacetime symmetries.} In particular for the coordinates chosen above 
this means that: 
\begin{equation}
\partial_t\Phi=0=\partial_\phi\Phi \ .
\end{equation}

\bigskip

Under these two assumptions, multiply the Klein-Gordon equation by $\Phi$ and integrate over the BH exterior space-time:
\begin{equation}
\int d^4x\sqrt{-g}\left[\Phi \nabla_\mu \nabla^\mu \Phi-\Phi V'\right]=0 \ .
\end{equation}
Now, integrating the first term by parts:
\begin{equation}
\int d^4x\sqrt{-g}\left[-\nabla_\mu\Phi \nabla^\mu \Phi-\Phi V'\right]+\int_{\mathcal{H}}d^3\sigma n^\mu\Phi \nabla_\mu\Phi=0 \ ,
\label{bek1}
\end{equation}
where the boundary term is computed on the horizon and the other boundary term (at infinity) vanishes since the scalar field should fall off sufficiently fast at infinity to guarantee asymptotic flatness (exponentially fast if there is a mass term).\footnote{In Hawking's version of this proof~\cite{Hawking:1972qk} the volume considered is also bounded ``in time" by two partial Cauchy surfaces. These give rise to two other surface terms upon the integration by parts, but these two terms precisely cancel one another.} 

The boundary term in  \eqref{bek1} is actually zero. Indeed, the event horizon of a stationary, asymptotically flat spacetime is a Killing horizon. Thus, the normal to $\mathcal{H}$, $n^\mu$, is a linear combination of the Killing vector fields; but the scalar field is invariant under these by Assumption 2.
 Thus $n^\mu \nabla_\mu\Phi=0$.
We conclude that\footnote{Here, we are implicitly assuming that $d^3\sigma$ and $\Phi$ are finite on $\mathcal{H}$.}
\begin{equation}
\int d^4x\sqrt{-g}\left\{\nabla_\mu\Phi \nabla^\mu \Phi+\Phi V'\right\}=0 \ .
\label{bek3}
\end{equation}

\bigskip

{\bf Assumption 3 (v.1): the potential $V$ obeys 
\begin{equation}
\Phi V'\ge 0 \ ,
\label{a4v1}
\end{equation}
everywhere, and $\Phi V'=0$ for (possibly) some discrete values $\Phi_i$.} Observe this holds for the aforementioned  non-self-interacting massive scalar field: $\Phi V'=\Phi^2\mu^2$. In Section~\ref{sec_drop1} we shall see a variation of this theorem leading to a different condition.

\bigskip

The gradient of $\Phi$ is orthogonal to both Killing vectors and thus must be spacelike or zero. Thus $\nabla_\mu\Phi \nabla^\mu \Phi\geqslant 0$. Then, since each term in the integrand of \eqref{bek3} is non-negative the equality holds iff $\Phi=0,\Phi_i$, which establishes the no-hair theorem (in the case $\Phi=\Phi_i$ the scalar field is a cosmological constant). Remarkably, this theorem, \textit{did not use} the Einstein equations.

%%%%%%%%%%%%%%%%%%%
\subsection{Further no-scalar-hair theorems}
\label{sec_32}
%%%%%%%%%%%%%%%%%%%
Violating one of the assumptions 1,2,3 does not guarantee, by itself, the existence of a regular BH with scalar hair as we now discuss. We shall address a set of further no-hair theorems that drop totally or in part one of these assumptions. Following an order of chronological development we shall address assumption 1,3, and then finally 2.

%%%%%%%%%%%%%%%%%%%
\subsubsection{Reconsidering assumption 1}
\label{sec_drop1}
%%%%%%%%%%%%%%%%%%%
Assumption 1 is violated, for instance, by scalar-tensor theories of gravity\cite{Damour:1992we}, of which the pioneering example is  Brans-Dicke theory~\cite{Brans:1961sx}. In this family of theories, there is a scalar field non-minimally coupled to the geometry, so that the scalar field equation involves the curvature. This scalar, $\varphi$, is part of the gravitational interaction, whereas the scalar in the previous sections, $\Phi$, is regarded as matter. But this distinction is conformal-frame dependent.

A no-scalar-hair theorem for scalar-tensor gravity was established by Hawking~\cite{Hawking:1972qk}, who showed that in the Brans-Dicke theory the regular BH solutions are the same as in general relativity. To establish this result, consider Brans-Dicke theory in the original Jordan frame, where it is described by the action:
\begin{equation}
\mathcal{S}_{\rm BD}^J=\int d^4x \sqrt{-\hat{g}}\left [ \frac{1}{16\pi}
\left(
\varphi\hat{R}-\frac{\omega_0}{\varphi}\hat{\nabla}_\mu\varphi\hat{\nabla}^\mu\varphi
\right)
+\mathcal{L}_m(\hat{g}_{\mu\nu},\Psi_m)\right] \ .
\end{equation}
The Brans-Dicke scalar $\varphi$, plays, physically, the role of a spacetime varying Newton's constant. Matter fields, 
here collectively denoted by $\Psi_m$, with matter Lagrangian density $\mathcal{L}_m$, couple minimally to the Brans-Dicke metric $\hat{g}$, which has Ricci scalar $\hat{R}$ and covariant derivative $\hat{\nabla}$. 
Thus, matter particles follow geodesics of this metric. Hawking's proof consisted on 
performing a conformal transformation of the metric $\hat{g}\rightarrow g$ 
(and for clarity we perform simultaneously a field redefinition $\varphi\rightarrow \Phi$): 
\begin{equation}
g_{\mu\nu}\equiv \varphi \hat{g}_{\mu\nu} \ , \qquad \Phi=\int \frac{d\varphi}{\varphi}\sqrt{\frac{2\omega_0 +3}{4}} \ .
\label{bdtransf}
\end{equation}
This yields the Brans-Dicke action in the Einstein frame, where the scalar field is minimally coupled to the conformally transformed metric: 
\begin{equation}
\mathcal{S}_{\rm BD}^E= \frac{1}{4\pi} \int d^4x \sqrt{-g}\left( \frac{R}{4}
-\frac{1}{2}\nabla_\mu\Phi \nabla^\mu\Phi +4 \pi e^{-\frac{4}{\sqrt{3+2\omega_0}}\Phi} \mathcal{L}_m\left(\frac{{g}_{\mu\nu}}{\varphi},\Psi_m\right)\right) \ .
\label{bd20}
\end{equation}
In this frame the locally measured Newton's constant is constant but the masses of particles vary as $\varphi^{-1/2}$ and thus massive particles do not move along geodesics. Moreover, the scalar field equation is the Klein-Gordon equation sourced by the trace of the matter energy--momentum tensor. Assuming space-time to be empty apart from electromagnetic fields -- which have vanishing energy-momentum trace -- the scalar field obeys the sourceless Klein-Gordon equation. Then, the action \eqref{bd20} reduces to \eqref{sv_action2} with $V(\Phi)=0$, up to the electromagnetic field, which does not change the scalar equation. Hawking applied precisely the argument of Section~\ref{sec_bek_th} to establish the Brans-Dicke scalar field must be zero outside the horizon. Since the scalar field is zero, the Brans-Dicke equations of motion reduce to those of general relativity and so must the BH solutions. Note that Hawking crucially assumed that the scalar field is invariant under the action of the Killing vector fields (Assumption 2 above). 

Hawking's theorem has been recently generalized to a more general class of scalar-tensor theories of gravity by Sotiriou and Faraoni.~\cite{Sotiriou:2011dz} Again the argument relies on a slight variation of the reasoning in Section~\ref{sec_bek_th}. The class of scalar-tensor theories considered is described by the Jordan frame action:
\begin{equation}
\mathcal{S}_{\rm ST}^J=\int d^4x \sqrt{-\hat{g}}\left[\frac{1}{16\pi}
\left(
\varphi\hat{R}-\frac{\omega(\varphi)}{\varphi}\hat{\nabla}_\mu\varphi\hat{\nabla}^\mu\varphi-U(\varphi)
\right)
+\mathcal{L}_m(\hat{g}_{\mu\nu},\Psi_m)\right] \ .
\label{gen_sca_ten}
\end{equation}
Thus, the Brans-Dicke scalar has now a potential, $U(\varphi)$ and the parameter $\omega_0$ becomes a function $\omega(\varphi)$. Nevertheless, applying the same transformations \eqref{bdtransf} as before (replacing $\omega_0\rightarrow \omega(\varphi)$) one obtains the Einstein frame action:
\begin{equation}
\mathcal{S}_{\rm ST}^E=\int d^4x \sqrt{-g}
\left[
\frac{1}{4\pi}
\left(
\frac{R}{4}-\frac{1}{2}\nabla_\mu\Phi \nabla^\mu\Phi-V(\Phi)
\right)
+\varphi(\Phi)^{-2}\mathcal{L}_m\left(\frac{{g}_{\mu\nu}}{\varphi},\Psi_m\right)
\right] \ ,
\label{bd2}
\end{equation}
where $V(\Phi)=U(\varphi)/\varphi^2$. For $\mathcal{L}_m=0$ one can immediately apply the argument of Section~\ref{sec_bek_th}, since \eqref{bd2} reduces to \eqref{sv_action2}. But one may devise a variation of this argument, which applies for $V(\Phi)\neq0$ and leads to a different condition on the potential energy. One considers the Klein-Gordon equation and, instead of multiplying it by $\Phi$, as in Section~\ref{sec_bek_th}, one multiplies by $V'(\Phi)$. Following the same steps one arrives at 
\begin{equation}
\int d^4x \sqrt{-g}\left[V''(\Phi)\nabla^\mu\Phi \nabla_\mu\Phi+[V'(\Phi)]^2\right]=0 \ .
\label{intsf}
\end{equation}
instead of \eqref{bek3}. The second term in the integrand is non-negative. Moreover, as before, 
under Assumption 2, $\nabla_\mu\Phi$, must be spacelike or zero, which implies that $\nabla_\mu\Phi \nabla^\mu\Phi\geqslant 0$. To finish addressing the first term, one considers the following assumption.

\bigskip

{\bf Assumption 3 (v.2): the potential energy is convex.} In other words:
\begin{equation}
V''(\Phi)\geqslant 0 \ .
\label{sotfar}
\end{equation}

\bigskip

Under this assumption, both terms in the integral \eqref{intsf} are non-negative. Thus, the scalar fields must be trivial in the exterior spacetime; hence the equations of motion of the scalar-tensor theory reduce to those of general relativity and so must the BH solutions.  Assumption~\eqref{sotfar} was given the interpretation of linear stability of the BH solution.~\cite{Sotiriou:2011dz}  As a word of caution, this means that this argument does not exclude unstable, but long lived, solutions, which could conceivably be physically relevant -- even though they do not violate the dynamical spirit of the no-hair theorem in what regards the very final end-state of gravitational collapse.  

How is this theorem compatible with the BBMB solution, eq.~\eqref{BBMB1}--\eqref{BBMB2}? Regarding conformal scalar-vacuum theory, described by \eqref{csv_action}, as a scalar-tensor theory of gravity with action~\eqref{gen_sca_ten}, we first observe there is no potential for the scalar field. Hence the obstruction is not related to \eqref{sotfar}.
 The scalar field $\varphi$ in~\eqref{gen_sca_ten} relates to the scalar field $\Phi$ in~\eqref{csv_action} as
\begin{equation}
\varphi=1-\frac{1}{3}\Phi^2 \ .
\end{equation} 
Substituting the explicit  BBMB solution for $\Phi$, eq.~\eqref{BBMB2}, one finds that $\varphi=r(r-2M)/(r-M)^2$. Thus $\varphi=0$ at $r=2M$ and the conformal transformation \eqref{bdtransf}  becomes singular at this point. This signals a breakdown of the conformal transformation given by~\eqref{bdtransf} and the inapplicability of the theorem.~\cite{Sotiriou:2011dz} Moreover, $\varphi<0$ for $M<r<2M$ and $\varphi\rightarrow \infty$ at $r=M$ (horizon). Thus, in this region, Newton's constant ``changes sign", presumably sourcing a type of anti-gravity which explains the existence of this hairy solution.\footnote{We thank C. Charmousis for this remark.}  As a final remark concerning conformal scalar-vacuum, Xanthopoulos and Zannias\cite{xanthopoulos:91} and Zannias~\cite{Zannias:1994jf}  established that the only static asymptotically flat non-extremal BH solution having the scalar field bounded on the horizon, is the Schwarzschild BH. Thus, moving away from extremality (of the BBMB geometry) does not allow hairy BHs with more independent parameters, and in particular no primary hair can be found. Other theorems covering non-minimal coupling were developed by Saa.\cite{Saa:1996aw,Saa:1996qq}

Further possible loopholes of the no-scalar-hair theorem for scalar-tensor theories are discussed in the original article,~\cite{Sotiriou:2011dz}  where the authors also invoke the weak energy condition ($cf.$ eq. \eqref{wec} below), so that stationarity implies axi-symmetry by Hawking's theorem.\cite{Hawking:1971vc}

\bigskip

Another class of theories where Assumption 1 is violated and which have been focus of recent interest are Horndeski\cite{Horndeski:1974wa} and, in particular, Galileon\cite{Nicolis:2008in}, or generalized Galileon \cite{Deffayet:2009mn} theories. Unlike the previously considered cases, these theories include second order derivatives of the scalar field in the action. 
Horndeski showed long ago -- a result recently rederived in the context of Galileon theories\cite{Deffayet:2011gz,Kobayashi:2011nu} -- that the most general scalar-tensor action with up to second order derivatives of the scalar field 
and with second order \textit{field equations} is given by
\begin{equation}
\begin{array}{l}
\mathcal{S}=\int d^4x \sqrt{-g}\left\{
K(\Phi,X)
-G_3(\Phi,X)\Box \Phi 
+G_4(\Phi,X)R
+G_{4X}\left[(\Box\Phi)^2-(\nabla_\mu \nabla_\nu\Phi)^2\right]  
   \right. \\ \\
\qquad   
\left.\displaystyle{
+G_{5}(\Phi,X)G_{\mu\nu}\nabla^\mu \nabla^\nu\Phi 
-\frac{G_{5X}}{6}\left[(\Box\Phi)^3-3\Box \Phi(\nabla_\mu \nabla_\nu\Phi)^2+2(\nabla_\mu \nabla_\nu\Phi)^3\right] }  \right\} \ ,
\end{array}
\label{ha}
\end{equation}
where $K,G_i$ are generic functions of $\Phi$, 
\begin{equation}
X\equiv -\frac{1}{2}\nabla_{\mu}\Phi \nabla^\mu\Phi \ ,
\label{x}
\end{equation} 
and $G_{iX}\equiv \partial G_i/\partial X$. The reason to focus on theories with at most second order field equations is that theories with higher order field equations are generically afflicted by Ostrogradski instabilities.\cite{ostro} In terms of the requirement on the order of the equations of motion, Horndeski theories are the scalar field analogue to Lovelock theories of vacuum gravity.\cite{Lovelock:1971yv}

Here we shall focus on Horndeski theories with shift symmetry, $i.e.$ invariant under $\Phi\rightarrow \Phi+$constant. The most general such theory is obtained from~\eqref{ha} simply by dropping the $\Phi$ dependence for $K,G_i$.\cite{Sotiriou:2014pfa} From the QFT viewpoint, this symmetry protects the scalar field from acquiring a mass term, due to radiative corrections.\cite{Sotiriou:2013qea} From the classical viewpoint, it implies the existence of a Noether current $J^\mu$, and the scalar field equation can be written as a conservation equation for this current, 
$\nabla_\mu J^\mu=0$. 

A no-scalar-hair theorem for static and spherically symmetric BHs (not necessarily asymptotically flat) in shift-symmetric Horndeski/Galileon theories was proposed by Hui and Nicolis.\cite{Hui:2012qt} We now sketch the argument. 
Writing the static and spherically symmetric line element in the gauge
\begin{equation}
\label{HN}
ds^2=-f(R)dt^2+\frac{dR^2}{f(R)}+r(R)^2(d\theta^2+\sin^2\theta d\phi^2) \ ,
\end{equation}
one starts assuming that the scalar field inherits the spacetime symmetries (Assumption 2 above); thus the scalar field depends only on the radial coordinate: $\Phi=\Phi(R)$. Moreover the 4-current only has the radial component $J^R$.
 Then 
 $J^\mu J_\mu=(J^R)^2/f$. 
 This is a physical quantity that should be well behaved at the horizon, where $f=0$. Thus, $J^R$ should vanish on the horizon. Integrating the scalar equation of motion  
 $\nabla_\mu J^\mu=0$ 
 yields $r(R)^2J^R=$constant. Since the areal radius $r(R)$ should be finite at a regular horizon, then $J^R=0$ everywhere. As a final step, it is argued that $J^R=0$ implies that $\Phi=$constant in the whole spacetime. This final point in the argument turns out to leave space for the existence of hairy solutions, $cf.$ Section \ref{sec_41}.

%%%%%%%%%%%%%%%%%%%
\subsubsection{Reconsidering assumption 3}
\label{sec_drop4}
%%%%%%%%%%%%%%%%%%%

Assumption 3, both v.1 and v.2, is violated by some physical potentials, like the Higgs potential.
A different type of no-scalar-hair theorem \cite{HS,Heusler:1996ft} for theories within the class (\ref{sv_action2}) 
that invoke the \textit{strong} energy condition use scaling techniques, $i.e.$ a 
curved space generalization of the original flat space Derrick-type argument \cite{Derrick:1964ww},
and apply to  spherically symmetric configurations. 
Let us take the generic spherically symmetric line element in the following form,
\begin{eqnarray}
\label{Scwh-like-metric}
ds^2=-N(r)\sigma^2(r) dt^2+\frac{dr^2}{N(r)}+r^2(d\theta^2+\sin^2\theta d\phi^2),~~~N(r)\equiv 1-\frac{2m(r)}{r}.
\label{schgen}
\end{eqnarray}
in terms of two unknown functions $m(r)$ and $\sigma(r)> 0$. The function $m(r)$ is related to the local mass-energy density inside a sphere of radius $r$, sometimes called Misner-Sharp mass function\cite{Misner:1964je}.
The BH horizon is located at $r=r_H>0$,
where $N(r_H)=0$ and $N'(r_H)\geq 0$. Thus, $2m(r_H)=r_H$.

Inserting the anstaz \eqref{schgen} into the action  (\ref{sv_action2}) and integrating the trivial angular dependence, one arrives at the 
effective action 
\begin{eqnarray}
\label{Seff}
S_{eff}=\int_{r_H}^\infty dr~
\sigma(r)
\bigg[
m'-\left( \frac{1}{2}N r^2 \Phi'^2+r^2 V(\Phi) \right)
\bigg] \ .
\end{eqnarray}
Let us now assume\cite{HS,Heusler:1996ft} the existence of a BH solution described by $ m(r)$, $\sigma(r)$ and $\Phi(r)$
 with suitable boundary conditions at the event horizon
$r=r_H$ and at infinity. 
 Then each member of the 1-parameter family
$
m_\lambda(r) \equiv m(r_H+\lambda (r-r_H)),
$
$
\sigma_{\lambda} (r) \equiv \sigma (r_H+\lambda (r-r_H)),
$
and
$
\Phi_{\lambda} (r) \equiv \Phi (r_H+\lambda (r-r_H)),
$
assumes the same boundary values at $r=r_H$ and at $r=\infty$, and the action 
$S_{\lambda} \equiv S[m_{\lambda}, \sigma_{\lambda}, \Phi_{\lambda}] $
must have a critical point at $\lambda=1$, $[dS/d\lambda]_{\lambda=1}=0$.
Thus any hairy BH solution must satisfy the  virial relation \cite{Heusler:1996ft}
\begin{eqnarray}
\label{virial1}
 \int_{r_H}^\infty dr~
\sigma(r)
\bigg[
\left( \frac{2r_H}{r}\left(1-\frac{m}{r}\right)-1
\right)
\frac{1}{2}r^2\Phi'^2+\left(\frac{2r_H}{r}-3\right)r^2V(\Phi)
\bigg]
=0 \ .
\end{eqnarray}
The prefactor of $V$ and the full first term are negative for $r\geqslant r_H$. 
Then one may evoke the positivity of  $V$ to establish the result.

\bigskip

{\bf Assumption 3 (v.3): the potential energy density is non-negative.} This requirement is
\begin{equation}
V(\Phi)\geqslant 0 \ .
\label{pospot}
\end{equation}

 \bigskip

For $V(\Phi)\geqslant 0$, the integrand in \eqref{virial1} is a negative
quantity. Thus,  regular BH solutions with non-trivial scalar hair solving the model (\ref{sv_action2}) must 
necessarily have a negative potential for some range of $r$.

\bigskip

Another class of  no-scalar-hair theorems make
direct use of the Einstein equations and as such focused on spherically symmetric line elements, for which the equations are considerably simpler than for the axi-symmetric case.\cite{Heusler:1992ss,Bekenstein:1995un,Sudarsky:1995zg}  
  In particular, Bekenstein could rule out spherically symmetric, static BHs, allowing the possibility of many scalar fields, possibly with non-canonical kinetic terms, either minimally coupled to gravity or with a specific non-minimal coupling (of Brans-Dicke type), assuming a non-negative energy density\cite{Bekenstein:1995un}, rather than Assumption 3 (v.1, v.2 or v.3): 

\bigskip

{\bf Assumption 3 (v.4): the energy density is non-negative ({\it weak energy condition}).} 
This must hold everywhere for any timelike observer. Then, the requirement is 
\begin{equation}
\rho\equiv T_{\mu\nu}U^\mu U^\nu\geqslant 0 \ .
\label{wec}
\end{equation}
For the energy-momentum tensor derived from \eqref{sv_action2}, 
 \begin{equation}
T_{\mu\nu}^S=\partial_\mu\Phi\partial_\nu\Phi-\frac{1}{2}g_{\mu\nu}\partial_\alpha\Phi\partial^\alpha\Phi-g_{\mu\nu}V \ ,
\label{tmnv}
 \end{equation}
for a static observer $U^\mu\propto \delta^\mu_t$ and under assumption 2 for a static spacetime, such that $\partial_t\Phi=0$, this requirement is
\begin{equation}
\rho=\frac{1}{2}\partial_\alpha \Phi \partial^\alpha \Phi+V\geqslant 0 \ .
\label{potweak}
\end{equation}

\bigskip

Notice that, in the context of spherically symmetric solutions with $\Phi=\Phi(r)$, a violation of the weak energy condition \eqref{potweak}, implies, generically, a violation of the strong energy condition \eqref{pospot}; the converse, however, is not true.

Bekenstein's ``novel no-hair theorem"\cite{Bekenstein:1995un} (see also Sudarsky's\cite{Sudarsky:1995zg}) is based on a careful analysis of the radial component of the scalar fields' energy--momentum tensor $T^r_{\ r}$ and its radial derivative $\partial_r T^r_{\ r}$. By using first the energy--momentum conservation equations and subsequently some of the Einstein equations, for a spherically symmetric line element, describing an asymptotically flat geometry with a regular horizon, a contradiction for the sign of $\partial_r T^r_{\ r}$ in the exterior spacetime is obtained, which can only be resolved if the scalar field is trivial outside the horizon. This argument has been generalized to higher dimensions.\cite{Skakala:2014gca}

\bigskip

As a final remark concerning different energy requirements, Hertog\cite{Hertog:2006rr} provided evidence for a no-scalar-hair theorem of relevance for string compactifications, ruling out spherical scalar hair of static BHs if the scalar field theory, when coupled to gravity, satisfies the Positive Energy Theorem.\cite{Witten:1981mf}
Another type of no-scalar-hair theorem using the strong energy condition 
is  based  on a mass bound for spherically symmetric BHs.\cite{Heusler:1994wa,Hb}

\bigskip

%%%%%%%%%%%%%%%%%%%
\subsubsection{Reconsidering assumption 2}
\label{nohairsym}
%%%%%%%%%%%%%%%%%%%
Assumption 2 has a different character compared to Assumptions 1,3, since it is not associated to changing the theory one is working with. Moreover, it seems quite natural -- almost obvious -- to assume that the scalar field has the same symmetries as the geometry. This is, however, not mandatory. What is mandatory, is that the energy-momentum tensor of the scalar field should share the symmetries of the geometry, which is not the same thing. The difference is illustrated by allowing the scalar field to be complex and possessing a harmonic time dependence, $\phi\sim e^{-iwt}$
(equivalently, one can consider $two$ real scalar fields with opposite phases and
the same mass). 
The complexity of the scalar field allows the energy momentum tensor to be time-independent, even though the scalar field is time dependent; hence it is compatible with static and spherically symmetric geometries. One explicit example are boson stars.\cite{Schunck:2003kk,Liebling:2012fv} These are self-gravitating, solitonic-like, scalar field configurations, first discussed long ago by Kaup\cite{Kaup:1968zz} and Ruffini and Bonazzola.\cite{Ruffini:1969qy} The scalar field must have a mass term and may or may not have self-interactions. 

It is reasonable to ask if the boson stars possess BH generalizations.
They would have scalar field hair with the same type of harmonic time dependence, 
 which renders this case outside the scope of previous no-scalar-hair theorems. 
 The answer, for spherically symmetric geometries, was provided by a no-hair theorem by Pena and Sudarsky.~\cite{Pena:1997cy} 
 This theorem established the inexistence of BH with non-trivial scalar field hair, considering one or more complex scalar fields minimally coupled to gravity that vary harmonically with time and with an arbitrary potential, subject to the condition that the energy-momentum tensor obeys the weak energy condition (Assumption 3 v.3), and that the radial pressure is no less than tangential pressures. The proof relies on a similar method to that of Bekenstein's ``novel no-hair-theorem" described in Section~\ref{sec_drop4}. In particular, the idea is to show that the harmonic time dependence changes the energy--momentum tensor into an effective one that still obeys the energy requirements.

\bigskip

Recently a theorem was reported by Graham and Jha that further rules out the existence of stationary, asymptotically flat (in particular) BH solutions of the Einstein equations coupled to a real, time dependent scalar field, for a class of scalar field actions.\cite{Graham:2014ina} Specifically, the theorem applies to theories with action:
\begin{equation}
\mathcal{S}=\frac{1}{4\pi}\int d^4x\sqrt{-g}\left(\frac{R}{4}+P(\Phi,X)\right) \ ,
\label{sv_action_gj}
\end{equation}
where $X$ is defined by eq. \eqref{x}. Thus, the action depends only on first derivates of the scalar field and not higher order ones, but it can have a non-canonical kinetic term. In particular, as an earlier theorem by the same authors\cite{Graham:2014mda}, it applies to the scalar fields used in $K-$essence models. The key point in the argument, based on an earlier observation\cite{Wyman:1981bd}, is that, assuming a stationary geometry  -- which is thus also axi-symmetric by the rigidity theorem, assuming the scalar field to obey the null energy condition -- and writing it in the form
\begin{equation}
ds^2=-e^{\mu(r,\theta)}dt^2+2\rho(r,\theta)dt d\phi+e^{\nu(r,\theta)}d\phi^2+e^{A(r,\theta)}dr^2+e^{B(r,\theta)}d\theta^2\ ,
\end{equation}
then, from the Einstein equations and the fact that $R_{tr}=0=R_{t\theta}$, it follows that
\begin{equation}
\partial_t\Phi\partial_r\Phi=0=\partial_t\Phi\partial_\theta\Phi \ .
\label{consca}
\end{equation}
Thus, assuming that $\Phi$ depends on $t$, it cannot depend on both $r$ and $\theta$ and thus $\Phi=\Phi(t,\phi)$. Then, inspection of the energy-momentum tensor and remaining Einstein equations reveals that $P=P(X)$, $i.e.$ no potential term is allowed, and $\Phi$ can only depend linearly on $t$ and $\phi$. The latter is excluded from the periodicity of $\phi$ and thus $\Phi=\alpha t+\beta$. Finally, asymptotic flatness requires, in general, that $\alpha=0$. We emphasize this theorem does not apply to more than one scalar field, or equivalently, to one (or more) complex scalar field. The reason is that, considering $\Phi$ complex, eq. \eqref{consca} is replaced by
\begin{equation}
\partial_{(t}\Phi^*\partial_{r)}\Phi=0=\partial_{(t}\Phi^*\partial_{\theta)}\Phi \ .
\end{equation}
Taking now, for instance, $\Phi=e^{-iwt}f(r,\theta)$, we see that this no longer constrains $f(r,\theta)$.

%%%%%%%%%%%%%%%%%%%%%%%%%%%%%%%%%%%%%%%%%%%%%%%%%%%
\subsubsection{Remarks}
%%%%%%%%%%%%%%%%%%%%%%%%%%%%%%%%%%%%%%%%%%%%%%%%%%%
Two remarks are in order concerning no-scalar-hair theorems for BHs and the hairy solutions presented in the next Section.  Firstly, we are focusing on independent scalar fields.
Considering \textit{simultaneously} gauge fields to which the scalar fields are non-minimally coupled leads to many different families of solutions. Notable examples include BHs and branes in supergravity~\cite{Gibbons:1982ih,Gibbons:1987ps,Horowitz:1991cd}, 
as well as early examples of BHs with scalar hair.\cite{Breitenlohner:1991aa,Greene:1992fw,Achucarro:1995nu}. 
All these have no independent scalar charge and the scalar field vanishes when the gauge field vanishes. Secondly, we shall be focusing on four dimensional (and asymptotically flat) BHs although some remarks are made concerning higher dimensional generalizations.

%%%%%%%%%%%%%%%%%%%
\section{Black Holes with scalar hair}
\label{sec_4}
%%%%%%%%%%%%%%%%%%%
We shall now describe explicit examples of asymptotically flat BHs with scalar hair (and no gauge fields) which are regular on and outside the event horizon. These solutions, of course, violate one of the assumptions discussed in Section \ref{sec_3}. The conceptually simplest route to obtain hairy solutions -- albeit physically questionable -- is to maintain a minimally coupled scalar field, with a canonical kinetic term and allow for a potential energy that violates the energy conditions discussed above (Assumption 3, v.1-4). We shall start by discussing this case. Next, we will consider solutions for non-minimally coupled scalar fields, thus violating Assumption 1. In this case the pioneering example is the BBMB BH already mentioned in Section~\ref{sec_23}. We shall further discuss BH solutions both  in the context of generalized scalar-tensor theories of the Horndeski/Galileon type and for scalar field theories with non-canonical kinetic terms. Finally, we shall address the most economic example of scalar hair, in what concerns the need of some exotic scalar field theory. This example occurs for a minimally coupled complex (or equivalently two real) scalar field, with a canonical kinetic term; $i.e.$ in Einstein-Klein-Gordon field theory with just a mass term. 
The corresponding {\it `Kerr BHs with scalar hair'} are made possible from the violation of Assumption 2 (and no others).

%%%%%%%%%%%%%%%%%%%
\subsection{Solutions violating assumption 3}
%%%%%%%%%%%%%%%%%%%

The simplest way to circumvent the no-scalar-hair theorems
-- and \textit{at first glance} the mildest deviation from the original model --
is to allow for a scalar field potential which is not strictly positive,
such that the strong (and generically also the weak) energy condition is violated.
Even though no clear physical settings for this violation appear to exist at this moment, 
at least in the asymptotically flat case, pragmatically assuming it leads to BHs with scalar hair possessing distinct properties.

Heuristically, a negative potential can provide an extra repulsive interaction which 
prevents the scalar field from infinitely piling up at the (would be) horizon -- as seen in the massless, non-self-interacting case -- when one requires $\Phi$ to be non-trivial. An attractive feature of some of these models is that a careful choice of
the expression for the scalar field potential
leads to exact solutions in closed form.

%%%%%%%%%%%%%%%%%%%%%%%%%%%%%%%%%%%%%%%%%%%%%%%%%%%%%%%%%%%%%%%%%%
\subsubsection{Scalar potential engineering: a test field solution}
%%%%%%%%%%%%%%%%%%%%%%%%%%%%%%%%%%%%%%%%%%%%%%%%%%%%%%%%%%%%%%%%%%
Consider once more the nonlinear Klein-Gordon
equation (\ref{n-KG}) for a test scalar field in the Schwarzschild BH background (\ref{sch}). 
Inspection of eq. \eqref{n-KG} reveals that \textit{choosing} a specific scalar radial profile $\Phi=\Phi(r)$ one obtains an equation of the form $V'=\chi(r)$, for some function $\chi(r)$. Inverting the radial profile $r=r(\Phi)$, one may choose the scalar potential to be
\begin{equation}
V(\Phi)=\int \chi(r(\Phi)) \, d\Phi \ .
\end{equation}
Thus, modulo subtleties in the inversion of $\Phi=\Phi(r)$, it is always possible to engineer a scalar
potential $V(\Phi)$ that allows any given scalar profile. 
In particular, one can construct smooth configurations 
with an energy-momentum tensor
which is
regular at the horizon and possess a finite total mass.

As perhaps the simplest example, let us choose the test scalar field to have the Coulomb-like form found for the electrostatic potential, $cf.$ eq. \eqref{pot_l}:
\begin{equation}
\Phi(r)=-\frac{Q_S}{r} \ .
\label{scalar-c}
\end{equation} 
This is a solution of (\ref{n-KG}) if one chooses the scalar potential to be
\begin{equation}
V(\Phi)=-\lambda \Phi^5<0 \ , 
\label{quintic}
\end{equation}
with $Q_S$ fixed by the coupling constant $\lambda>0$ and the BH mass $M$ 
\begin{equation}
\label{Qs}
Q_S=-\left(\frac{2M}{5 \lambda}\right)^{1/3}<0 \ .
\end{equation}
Observe that $Q_S\rightarrow \infty$ as $\lambda\rightarrow 0$. For any finite $\lambda$, the energy-momentum tensor of this test scalar field is 
finite on the horizon; the time-time component and the finite total mass-energy are 
\begin{equation}
(T^S)_t^t=\frac{14M-5r}{2^{1/3}5^{5/3}r^5}\left(\frac{M}{\lambda} \right)^{2/3}, \ \qquad 
E=\frac{3\pi}{2^{1/3}5^{5/3}} ( M \lambda^2 )^{-1/3} \ .
\end{equation}

%%%%%%%%%%%%%%%%%%%%%%%%%%%%%%%%%%%%%%%%%%%%%%%%%%%%%%%%%%%%%%%%%%%%%%%%%
 \subsubsection{Beyond probe limit: backreacting solutions}
%%%%%%%%%%%%%%%%%%%%%%%%%%%%%%%%%%%%%%%%%%%%%%%%%%%%%%%%%%%%%%%%%%%%%%%%%

The simple example above with the quintic potential \eqref{quintic} captures a number of basic 
features of non-linear hairy BH solutions in $V$-scalar-vacuum. 
In fact, this test field configuration can be promoted into an exact solution of the full theory. As an example, insisting in keeping the $1/r$ scalar profile everywhere, in terms of the areal radius,
the scalar field potential must get corrected in the non-linear solution and gains the more complicated form\footnote{To the best of our knowledge the solutions 
(\ref{new-pot})-(\ref{new-sol})
have not been discussed in the literature previously; 
a Coulombic scalar profile has been considered in an exact solution\cite{Dennhardt:1996cz} but not in terms of the areal radius.}
\begin{equation}
\label{new-pot}
V(\Phi)=-\frac{15}{2}\lambda e^{\frac{\Phi^2}{2}}W+\frac{1}{2 Q_S^2}
\left(
(1+2e^{ \Phi^2})( \Phi^2-3)+\frac{W^2-27}{ \Phi^2-3 }
\right)\ ,~~
\end{equation}
with
\begin{equation}
W\equiv 3 \Phi+ \sqrt{\frac{\pi}{2}} e^{\frac{\Phi^2}{2}}( \Phi^2-3) {\rm Erf}(\frac{\Phi}{\sqrt{2}}) \ .
\end{equation}
The metric functions which enter the line element -- of the form  (\ref{Scwh-like-metric}) -- are given by
\begin{eqnarray}
\label{new-sol}
&&
m(r)=\frac{r^3}{ Q_S^2}
\bigg[
1+\frac{ Q_S^2}{r^2}
+e^{\frac{ Q_S^2}{r^2}}( \frac{225}{8}\lambda^2 Q_S^4-1)
\\
\nonumber
&&{~~~~~}
-\frac{1}{2}e^{\frac {Q_S^2}{ r^2}} 
\bigg(
\frac{15}{2} \lambda Q_S^2
-\frac{Q_S}{r}e^{-\frac{Q_S^2}{2r^2}}
+\sqrt{\frac{\pi}{2}}{\rm Erf}(\frac{Q_S}{\sqrt{2}r})
\bigg)^2
\bigg], ~~~ 
\sigma(r)=e^{-\frac{Q_S^2}{2r^2}} \ .
\end{eqnarray}This solution represents a BH for some range of the free parameters  $\lambda,Q_S$,
with a regular horizon of spherical topology.
Moreover, the scalar field and its energy density 
are finite everywhere (in particular on the horizon) 
and decay fast enough such that
the total mass-energy is finite.
Surprisingly, the  relation  (\ref{Qs}) between  $Q_S,\lambda$ and $M$
still holds at the fully nonlinear level. In particular this means the scalar hair here is of secondary type, as there is no independent scalar charge; it is totally determined by the BH mass and the coupling in the Lagrangian. This property seems to be common to models of this sort.

We remark that, recently, an asymptotically flat BH in closed form, 
with a regular horizon and scalar hair with the \textit{asymptotic} scalar field profile 
(\ref{scalar-c}) 
was found by allowing the potential to have a more generic form, 
but reducing \textit{asymptotically} to the quintic potential \eqref{quintic}.\cite{Cadoni:2015gfa}
In fact, there are  several other similar examples of 
 closed form solutions with scalar hair,
which were constructed in closed form, by carefully engineering 
the expression of the scalar field potential 
\cite{Bechmann:1995sa,Dennhardt:1996cz,Bronnikov:2001ah,Zloshchastiev:2004ny,Anabalon:2013qua,Anabalon:2012ih}.

By contrast, solutions with simpler and better motivated scalar field potentials 
can be found only by using numerical methods 
\cite{Nucamendi:1995ex,Gubser:2005ih,Corichi:2005pa,Kleihaus:2013tba}. 
For instance a model by Nucamendi and Salgado \cite{Nucamendi:1995ex} uses the potential
\begin{equation}
V(\Phi)=\frac{\lambda}{4}\left[(\Phi-a)^2-\frac{4(\eta_1+\eta_2)}{3}(\Phi-a)+2\eta_1\eta_2\right](\Phi-a)^2 \ ,
\end{equation}
where $\lambda,\eta_i,a$ are constants chosen such that the potential violates Assumption 3 v.3 
-- there is a global minimum which is negative -- 
and a local minimum at which $V=0$; the field tends asymptotically to this local minimum to guarantee asymptotic flatness.

Given the ambiguity in the choice of the  scalar potential,
it is rather difficult to identify  generic properties of this class of hairy
BHs.
However, these configurations
appear to be unstable against linear fluctuations 
\cite{Anabalon:2013baa,Nucamendi:1995ex,Kleihaus:2013tba}
(see, however  \cite{Dennhardt:1996cz}).
Another interesting property of 
most of these solutions is that they do not trivialize in the limit of vanishing event horizon; the limiting configurations
describe globally
regular, particle-like objects, known as $scalarons$\cite{Nucamendi:1995ex}. Although no generalizations of the Kerr BH were found so far within this class of models, there are, however, slowly rotating solutions\cite{Anabalon:2014lea}.
Also, a set of static axially symmetric deformations of
the Schwarzschild BHs were reported\cite{Kleihaus:2013tba},
for a model with a complex scalar field violating also Assumption 2.

\bigskip

Finally, let us mention another class of BHs with scalar hair which violates assumption 3:  {\it  phantom BHs}.
Such solutions exist in models
with a `plus' sign for the scalar field kinetic term in  the action (\ref{sv_action2}).
By itself, this change in sign does not suffice to yield regular BH solutions. Indeed, for the model \eqref{sv_action} changing only the sign of the scalar kinetic term,  the
Fisher-Janis-Newman-Winicour configuration 
(\ref{arealradius})-(\ref{jnw})
still solves the field equations,
this time with
$\mu= \sqrt{1-\frac{Q_S^2}{M^2}} $.
In order to find 
regular configurations within the theory  (\ref{sv_action2}), one needs, therefore, to turn on the scalar field potential.
This potential, however, is still subject to
a number of constraints, as a result of Bekenstein's theorem and its various refinements
mentioned above.
For example, $V(\Phi)$ must have both 
 both positive and negative values. 
 Solutions of this type have been constructed\cite{Bronnikov:2005gm,Dzhunushaliev:2008bq,Bronnikov:2012ch}, but are expected to be, generically, unstable.

%%%%%%%%%%%%%%%%%%%
\subsection{Solutions violating assumption 1}
\label{sec_41}
%%%%%%%%%%%%%%%%%%%
Allowing for more general scalar-tensor gravity theories than $V$-scalar-vacuum, different setups can lead to scalar-hairy BHs. The pioneering example was found within conformal-scalar-vacuum, corresponding to the BBMB BH, which has been generalized (in particular) to include rotation.\cite{Astorino:2014mda} In the following we shall discuss examples with non-canonical kinetic terms and more general non-minimal couplings between the scalar field and gravity.

%%%%%%%%%%%%%%%%%%%
\subsubsection{Horndeski/Galileon theories}
\label{horgal}
%%%%%%%%%%%%%%%%%%%

BH solutions with a non-trivial scalar field have been the subject of recent interest in the context of Horndeski/shift symmetric generalized Galileon theories. Loopholes in the no-hair theorem of Hui and Nicolis\cite{Hui:2012qt}, $cf.$ Section~\ref{sec_drop1}, were pointed out by two different groups\cite{Sotiriou:2013qea,Sotiriou:2014pfa,Babichev:2013cya}. These authors constructed explicit solutions using different strategies.

Babichev and Charmousis  considered the following action:\cite{Babichev:2013cya}
\begin{equation}
\label{BC}
\mathcal{S}=\int d^4x \sqrt{-g}\left[\zeta R-\eta \nabla_\mu \Phi \nabla^\mu \Phi+\beta G^{\mu\nu}\nabla_\mu \Phi \nabla_\nu \Phi -2\Lambda\right] \ ,
\end{equation}
where $G^{\mu\nu}$ is the Einstein tensor, $\zeta>0$, $\eta$ and $\beta$ are constants and $\Lambda$ is the cosmological constant. The way these authors circumvent the no-hair theorem of Hui and Nicolis is related to the last step in their argument. For this theory the current is $J^{\mu}=(\eta g^{\mu\nu}-\beta G^{\mu\nu})\partial_\nu\Phi$. Thus eliminating the radial component of the current can be achieved by imposing $\eta g^{\mu\nu}=\beta G^{\mu\nu}$, rather than $\Phi=$constant. Then Babichev and Charmousis observed that regular solutions could be found if one allows the scalar field to have a \textit{linear time dependence} (thus violationg also Assumption 2). This is consistent with requiring the BH spacetime to be spherical and static, since the scalar field only enters the equations of motion through its derivatives. Using  these methods, these authors constructed, in particular, a solution with precisely the Schwarzschild geometry  (\ref{sch})~Êand a non trivial scalar field:
\begin{equation}
\Phi=qt\pm 2qM\left[2\sqrt{\frac{r}{2M}}+\log \frac{\sqrt{r}-\sqrt{2M}}{\sqrt{r}+\sqrt{2M}}\right]+\Phi_0 \ ,
\label{BC2}
\end{equation}
where $q,\Phi_0$ are constants. For a non-trivial scalar field it is essential that $q\neq 0$. But in this case note that $\Phi$ is invariant under $q\rightarrow q/\beta$, $t\rightarrow \beta t$, $r\rightarrow \beta r$,$M\rightarrow \beta M$, whereas the metric scales by $\beta^2$. It should be noted that, depending on the chosen sign in the previous equation, the scalar field will diverge at either the future or past event horizon. Still, the geometry is regular on and outside the horizon. This construction, dubbed ``dressing a BH with a time dependent Galileon" has been applied to other shift symmetric subclasses of the Horndeski theory,\cite{Kobayashi:2014eva} bi-scalar extensions of Horndeski theory,\cite{Charmousis:2014zaa} $f(R)$ gravity\cite{Zhong:2015ina} and other examples.\cite{Charmousis:2015aya,Babichev:2015rva} Other static BH solutions in an extension of the model (\ref{BC})
with a scalar field mass have also been discussed. 
\cite{Kolyvaris:2011fk,Kolyvaris:2013zfa,Anabalon:2013oea}
Finally, observe the similarity between~\eqref{BC2} and~\eqref{jacobson}. But whereas the latter is a test field solution on the Schwarzschild geometry for scalar-vacuum Einstein's gravity~\eqref{sv_action}, the former, together with the same geometry, forms an exact solution of the model~\eqref{BC}.

On the other hand,
Sotiriou and Zhou considered the following action:\cite{Sotiriou:2014pfa}
\begin{equation}
\mathcal{S}=\frac{1}{4\pi}\int d^4x\sqrt{-g}\left(\frac{R}{4}-\frac{1}{2}\nabla_\mu\Phi \nabla^\mu\Phi+\alpha \Phi\mathcal{G}\right) \ ,
\label{gb}
\end{equation}
where $\mathcal{G}$ is the Gauss-Bonnet combination
\begin{equation}
\mathcal{G}=R^{\mu\nu\rho\sigma}R_{\mu\nu\rho\sigma}-4R_{\mu\nu}R^{\mu\nu}+R^2\ .
\end{equation}
It is not immediately obvious that this action is a special case of the general action~\eqref{ha}. 
But indeed this action corresponds to the choice 
$K=G_3=G_4=0$ and $G_5=-4\alpha \ln |X|$.\cite{Kobayashi:2011nu} For this type of theory the last step of the argument in the no-hair theorem of Hui and Nicolis is circumvented, $cf.$ Section~\ref{sec_drop1}; actually, general solutions 
of this theory, including BH solutions, \textit{must} have a non-trivial scalar field.\cite{Sotiriou:2014pfa}
 Spherically symmetric BH solutions were constructed numerically 
 and have the interesting property of possessing a minimum size, controlled by the Gauss-Bonnet coupling constant $\alpha$.\cite{Sotiriou:2014pfa}
 
We remark that the first example of this subsection actually defines a method that can be used in obtaining explicit and analytic hairy solutions. Another example found in a bi-scalar extension of Horndeski theory\cite{Charmousis:2014zaa} corresponds to a non-extremal or extremal Reissner-Nordstr\"om geometry with \textit{primary} scalar hair, $cf.$ eqs. (5.24)--(5.27) therein.  By contrast, the second example is more specific and can only be tackled numerically.  Also, and in agreement with the discussion in the next subsection, the hair is secondary in this example. 
%%%%%%%%%%%%%%%%%%%%%%%%%%%%%%%%%%%%%%%%%%%%%%%%%%%%%%%%%%%%%%%%%%%%%%%%%%
\subsubsection{Scalar fields coupled to higher order curvature terms}
\label{sec_hc}
%%%%%%%%%%%%%%%%%%%%%%%%%%%%%%%%%%%%%%%%%%%%%%%%%%%%%%%%%%%%%%%%%%%%%%%%%%%%%%%%%%%%
The action (\ref{gb}) also belongs to a more general class of theories
where the scalar field couples non-minimally 
to higher order curvature terms.
For instance, in string gravity, wherein the scalar field is dubbed ``dilaton", 
one encounters an action of the form \eqref{gb} but with $\Phi\mathcal{G}$ replaced by $e^{\Phi}\mathcal{G}$. Thus, in four spacetime dimension where $\mathcal{G}$ is topological, the former can be regarded as a linearized version (in $\Phi$) of the latter.

The action of a general theory of gravity which includes all quadratic,
algebraic curvature invariants, generically coupled to a
single scalar field would contain in addition to (\ref{sv_action2}) 
the following terms\cite{Yunes:2011we},
\begin{eqnarray}
\label{sup}
&&\mathcal{S}=\int d^4 x\sqrt{-g} \bigg(
f_1(\Phi )R^2
+f_2(\Phi )R_{\mu\nu}R^{\mu\nu}
\\
\nonumber
&&{~~~~~~~~~~~~}
+f_3(\Phi )R_{\mu\nu\rho\sigma}R^{\mu\nu\rho\sigma}
+f_4(\Phi )R_{\mu\nu\rho\sigma} {\,^\ast\!} R^{\mu\nu\rho\sigma}
\bigg) \ ,
\end{eqnarray}
where ${\,^\ast\!} R_{\mu\nu\rho\sigma}$ is the dual Riemann tensor.
Theories of this type
are motivated from fundamental physics, such as in low-energy
expansions of string theory.
Also, in most studies,
the scalar field has no potential. Heuristically, the existence of BHs with scalar hair for such models
can be traced back again to the occurrence of \textit{effective} negative energy densities.
Indeed, the modified Einstein equations leads to an effective energy-momentum tensor that involves 
a supplementary
contribution from  (\ref{sup}).
This effective energy-momentum tensor may violate the weak energy condition.

The existence of these new couplings between the scalar field and curvature leads to source terms in the Klein-Gordon equation,
which may circumvent the various no-hair theorems.
The most studied case in this context corresponds to
the Einstein-Gauss-Bonnet-dilaton model, with
$f_1=\alpha e^{\gamma \Phi}$,
$f_2= -4 f_1$
and
$f_3= f_1$.
Although the BHs in this theory cannot be
found in analytical form\footnote{Note that these studies usually 
consider a more general matter content, with extra-gauge fields
coupled to dilaton, which, however, are not required for the existence
of hairy solutions.}, the static solutions were 
studied perturbatively in the
small coupling limit
\cite{Mignemi:1992nt,Mignemi:1993ce} 
and numerically for general
coupling\cite{Kanti:1995vq,Torii:1996yi,Alexeev:1996vs}.
The scalar hair is, however, of secondary type, as it is not independent from the BH mass.
The Kerr BH with dilatonic
Gauss-Bonnet corrections was considered numerically for slow rotation,\cite{Ayzenberg:2014aka} whereas its highly spinning counterpart was  constructed by Kleihaus et al.\cite{Kleihaus:2011tg} Slowly rotating solutions to~\eqref{sup} taking $f_i$ to be linear functions of $\Phi$  have been considered by Pani et al.\cite{Pani:2011gy}

Such solutions have 
many similar properties to those of their general relativity counterparts;  for example, the spherically symmetric solutions were claimed to be linearly stable.\cite{Kanti:1997br,Pani:2009wy}
There are however some novel properties. For example, 
some spinning BH solutions to Einstein-Gauss-Bonet-dilaton gravity may violate the Kerr bound.\cite{Kleihaus:2011tg} Also, in contrast to some other cases of scalar hairy BHs, 
they do not possess a solitonic limit.

Another situation of interest has $f_1=f_2=f_2=0$ and $f_4=\alpha \Phi$,
in which case 
the  theory (\ref{sup}) 
reduces to Chern-Simons gravity \cite{Alexander:2009tp}.
The  solutions of this model would differ from those of general relativity in the spinning case only.
However, only perturbative BHs have been constructed so far
\cite{Yunes:2009hc, Stein:2014xba}.  

Further properties of the  hairy BHs  induced by the  general action (\ref{sup})
with generic values of $f_i$
are discussed in the recent review by Berti et al.\cite{Berti:2015itd} Let us also mention that
BHs in theories with a scalar field coupled to even higher curvature terms (quartic) have been considered by Myers.\cite{Myers:1987qx}

%%%%%%%%%%%%%%%%%%%%%%%%%%%%%%%%%%%%%%%%%%%%%%%%%%%%%%%%%%%%%%%%%%%%%%%%%%%%%%
\subsubsection{Scalar field multiplets and non-canonical kinetic terms}
%%%%%%%%%%%%%%%%%%%%%%%%%%%%%%%%%%%%%%%%%%%%%%%%%%%%%%%%%%%%%%%%%%%%%%%%%%%%%%
As already mentioned generalization of the setting in Section 2 is to allow for multiple
scalar fields, eventually subject to some constraint.
Moreover, the scalar fields Lagrangian may posses a global $O(n)$ symmetry. 
A further ingredient is to allow for non-canonical kinetic terms for the scalar fields.
This makes possible the existence of localized, globally regular, finite energy field theory solutions 
already in the flat spacetime limit.
Such configurations have found important physical applications in various contexts ranging
from the models of condensed matter physics  to high energy physics and cosmology, as discussed in various reviews.\cite{Rajaraman,Manton,Radu:2008pp}

Self-gravitating lumps 
based on such Lagrangians minimally coupled do Einstein's gravity
 exist as well, and can be further generalized by
replacing the regular center by a black hole with a small horizon radius.
As such, in contrast to most of the
solutions in Sections \ref{horgal}, \ref{sec_hc},
this class of BHs follow the paradigm of {\it ``event horizons inside classical lumps"}
\cite{Kastor:1992qy}.

A simple example of BH solutions with scalar hair in this context 
is found in a model featuring an $O(3)$ scalar isovector field,
with the action
\begin{eqnarray}
\label{action-O3}
\mathcal{S}=\frac{1}{4\pi}\int d^4x\sqrt{-g}
\left(
\frac{R}{4}-\frac{1}{2}\nabla_\mu \Phi^a \nabla^\mu \Phi^a
-\frac{\lambda}{4}(\Phi^a \Phi^a-v^2)
\right) \ ,
\end{eqnarray}
where $v$ and $\lambda$ are the vacuum expectation value and
the self-coupling constant of the scalar fields, respectively.
The spherically symmetric solutions  have 
 a scalar field ansatz\footnote{Note that this leads to spherically symmetric configurations, such that the Assumption 2 
is violated; this applies as well for Skyrme hair.}
\begin{eqnarray}
\label{SO3}
\Phi^1=f(r)\sin\theta \cos\phi,~~\Phi^2=f(r)\sin\theta \sin\phi,~~\Phi^3=f(r)\cos\theta \ ,
\end{eqnarray}
(where $f(r)\to v$ as $r\to \infty$),
and a line element given by (\ref{Scwh-like-metric}).
 These are BHs inside global monopoles\cite{Barriola:1989hx}, which were discussed by Liebling\cite{Liebling:1999ke} and Maison\cite{Maison:1999ke}.
In contrast to, for instance, the BBMB case, such solutions -- including the scalar fields -- are regular on and outside the horizon.
However, their energy density  decays too slowly at
large distances.
Consequently, their mass, as defined in the usual way, diverges, 
even though a proper ADM mass can still be defined\cite{Nucamendi:1996ac,Nucamendi:2000af}. 
This leads to a deficit solid angle in the geometry of the space and the resulting spacetime is not strictly
asymptotically flat.
In fact, a number of no-hair theorems could be generalized \cite{Heusler:1996ft,Hb}
for the case of scalar multiplets,
by assuming the kinetic term of the scalar fields to be of the form
$G_{ab}(\Phi)\nabla_\mu \Phi^a\nabla^\mu \Phi^a$
(with $G_{ab}>0$).

The situation changes when allowing  
for non-standard kinetic terms of the scalar fields
in addition to the usual one.
Such Lagrangians are used 
in flat space field theory models 
\cite{Rajaraman,Manton,Radu:2008pp}
to evade 
Derrick-type scaling arguments\cite{Derrick:1964ww}, which forbid the existence of static scalar solitons.
Physically, the repulsive force associated with the extra-term allows for bound states.

The best known example in this context 
stems from the flat space Skyrme model 
\cite{Skyrme:1961vq,Skyrme:1962vh},
which can be regarded as an effective theory in the
low energy limit of QCD.
This model contains four scalars $\Phi^a$ 
satisfying the sigma-model constraint 
$\Phi^a\Phi^a=1$,
with a Lagrangian density
featuring a global $O(4)$ symmetry.
The action of the  \textit{Einstein-Skyrme model} can be written as\footnote{Here we use a slightly different formulation of the Skyrme model
as compared to the standard one in the literature in terms of an SU(2)-valued matrix $U$.}  
\begin{eqnarray}
\label{action-Skyrme}
\mathcal{S}=\frac{1}{4\pi}\int d^4x\sqrt{-g}
\left(
\frac{R}{4}
-\frac{1}{2}\nabla_\mu \Phi^a \nabla^\mu \Phi^a 
-\kappa
|\nabla_{[\mu} \Phi^a \nabla_{\nu]} \Phi^b|^2 
\right)\  ,
\end{eqnarray}
with $\kappa>0$.
This model has been studied by various authors,
mainly for the spherically case, 
for a scalar ansatz
\begin{eqnarray}
\Phi^1=\sin\chi(r) \sin\theta \cos\phi,~\Phi^2=\sin\chi(r) \sin\theta \sin\phi,~\Phi^3=\sin\chi(r) \cos\theta ,~\Phi^4=\cos\chi(r),
\nonumber
\end{eqnarray}
(with $\chi(r)$ the scalar amplitude),
 and
  the line element (\ref{Scwh-like-metric}).
The corresponding BH solutions -- \textit{BHs with Skyrme hair} -- were first discussed as test field Skyrmions around a Schwarzschild BH\cite{Luckock:1986tr},
and later including the backreaction \cite{Luckock}.
Argueably, these provide the 
first physically relevant counter-example to the no-hair conjecture
in the literature.
Further studies of their properties followed.
 \cite{Droz:1991cx, Heusler:1991xx,Heusler:1992av,Bizon:1992gb,Heusler:1993ci,Torii:1993vm}
In contrast to the BHs in the previous sections, the  hair of
the Skyrme BHs is primary, possesing a topological origin.

Interestingly, the BHs with Skyrme hair have a number of
features generic to certain models involving non-Abelian gauge fields,
with an intrincate branch structure.
Similar to other models possessing
flat space solitons,
the limit of zero horizon size 
corresponds to the smooth particle-like gravitating Skyrmions.
Moreover, one finds stable hairy configurations
against spherical linear perturbations. 
A review of these solutions in a more general context is provided by Volkov and Gal'tsov.\cite{Volkov:1998cc}

It is interesting to note that, in contrast to the models discussed in Section~\ref{sec_2}, $cf.$ Section~\ref{nss}, 
the Einstein-Skyrme model has BH solutions which are static, have a regular horizon
and possess  axially symmetry only.\cite{Shiiki:2005pb,Sawado:2003at}
A further comment is that no rotating BHs with Skyrme hair have been constructed so far, 
even though they should exist and their solitonic limit has been studied.\cite{Ioannidou:2006nn}

An interesting extension of this type of solutions 
is found when relaxing the sigma-model constraint
$\Phi^a \Phi^a=1$, imposing it only asymptotically.
 For example, the total mass of the BH solutions of the $O(3)$ scalar isovector field
 model discussed above is regularized
for an action with  Skyrme-like higher derivative terms.\cite{Radu:2011uj} 
This leads to asymptotically flat BHs
 with scalar hair which are regular on and outside the horizon, with
 the scalar field  asymptotically approaching a non-zero vacuum expectation value.
 Moreover, some of these solutions are stable  against time-dependent 
 linear fluctuations. 
Such BHs possess a regular extremal BH limit,
despite the absence of a scalar charge
associated with a Gauss law.

%%%%%%%%%%%%%%%%%%%
\subsection{Solutions violating assumption 2: Kerr BHs with scalar hair}
\label{kbhsh}
%%%%%%%%%%%%%%%%%%%
The solutions in the previous two subsections involve endowing the scalar field with some non-trivial dynamics due to either some (often unphysical) self-interactions or some modification of its propagator/coupling to gravity. Are there, however, scalar-hairy BHs in a theory with canonical kinetic terms, minimal coupling to gravity (Assumption 1) and abiding all energy conditions (Assumptions 3)? The only possibility is then to violate assumption 2.
 Such BHs indeed exist as we shall now describe, 
 and have a clear physical motivation.\cite{Herdeiro:2014goa,Herdeiro:2014ima,Herdeiro:2015gia}

We consider a complex scalar field $\Phi$, with a mass $\mu$, minimally coupled to Einstein's gravity. The corresponding action is:
\begin{equation}
\mathcal{S}= 
\frac{1}{4\pi}\int  d^4x\sqrt{-g}\left[\frac{R }{4}
   -\nabla_\mu \Phi^* \nabla^\mu\Phi - \mu^2 \Phi^*\Phi
 \right] \ .
 \label{complex}
 \end{equation}
It is again useful to start with a test field analysis. Linearizing the field equations on the scalar field one must solve the massive Klein-Gordon equation $\Box \Phi=\mu^2\Phi$ around a vacuum solution of the Einstein equations $R_{\mu\nu}=0$. We take the latter to be the Kerr solution in Boyer-Lindquist coordinates $(t,r,\theta,\phi)$ as, by the uniqueness theorems, this is the most general regular (on and outside an event horizon), asymptotically flat BH solution of vacuum general relativity. Next, in order to violate Assumption 2, we take the scalar field to depend on $t,\phi$. Since our final goal is to find some non-linear solution of \eqref{complex} we must guarantee that the energy momentum tensor of the scalar field is compatible with the background symmetries of a stationary (and thus axi-symmetric) BH. The presence of the mass term rules out, for instance, the linear time dependence considered in Sections \ref{nohairsym} and \ref{horgal}. But one may take a harmonic $t$ and $\phi$ dependence. Thus the scalar field ansatz is
\begin{equation}
\Phi=e^{-i w t} e^{im\phi} S_{\ell m} (\theta)R_{\ell m} (r) \ ,
\end{equation}
 where $w$ is the frequency, $m\in \mathbb{Z}$ and $S_{\ell m}$ are the spheroidal harmonics, $-\ell\leq m\leq \ell$, which can be defined as solutions of a second order ODE (for $\theta$) and $R_{\ell m}$ obey a second order ODE (for $r$).\cite{Brill:1972xj} This type of ansatz is reminiscent of that used for stationary states in quantum mechanics, which have a real frequency $w$. In particular the mass term creates a potential barrier at infinity which, in principle, can allow for bound states with an exponential fall-off towards infinity. 
 
 For a scalar field on a BH background, however, one must impose a purely ingoing boundary condition on the horizon (in a co-rotating frame). Thus one expects to find only solutions with \textit{complex} frequency $w=w_R+iw_I$ and $w_I<0$ signaling a decay of the scalar field towards the horizon. This is indeed the only type of solutions found in the Schwarzschild case (wherein $e^{im\phi} S_{\ell m}\rightarrow Y_\ell^m(\theta,\phi)$); such states are dubbed \textit{quasi-bound states}\footnote{These states can, however, be very long lived.\cite{Barranco:2012qs,Sanchis-Gual:2014ewa}}, which evolve in time.\cite{Burt:2011pv,Witek:2012tr,Okawa:2014nda} In the Kerr case, however, this only occurs for $w>m\Omega_H$, where $\Omega_H$ is the angular velocity of the Kerr horizon. For $w<m\Omega_H$ it turns out that $w_I>0$ and the amplitude of the scalar field \textit{grows} in time. This is due to the phenomenon of \textit{superradiance}\cite{Brito:2015oca}: the scalar field can be amplified in a scattering process with a Kerr BH, by virtue of extracting rotational energy from the BH. For bound states multiple superradiant scattering triggers the BH bomb as discussed by Press and Teukolsky.\cite{Press:1972zz} At the threshold of superradiance, $i.e.$ when
 \begin{equation}
 w=m\Omega_H \ ,
 \label{sync}
 \end{equation}
Hod observed that there are \textit{bound states} with real frequency.\cite{Hod:2012px,Hod:2013zza} These scalar bound states are regular on and outside the horizon and, due to the complex nature of the scalar field, they source a $t,\phi$-independent energy-momentum tensor. These have been called \textit{scalar clouds} around Kerr BHs.\cite{Hod:2012px,Hod:2013zza,Herdeiro:2014goa,Hod:2014baa,Benone:2014ssa} 

Scalar clouds can be promoted to a non-linear solution of the full massive--complex--scalar-vacuum system, much in the same way as the analysis of the Maxwell field in Section \ref{sec_21} leads to the existence of the Reissner-Nordstr\"om solution. A crucial aspect in this scalar case, however, is that the requirement of a real and time independent energy-momentum tensor demands considering a model with a $complex$ scalar field, or equivalently, two real fields with the same mass, and an $U(1)=O(2)$ global symmetry. The existence of these non-linear solutions was established, by numerical methods and the corresponding solutions dubbed \textit{Kerr BHs with scalar hair}.\cite{Herdeiro:2014goa,Herdeiro:2014ima,Herdeiro:2015gia} These are asymptotically flat, regular on and outside an event horizon rotating BH solutions with \textit{primary} scalar hair. Indeed, there is an independent charge counting the amount of scalar hair, a Noether charge associated to the $U(1)$ global invariance of the action \eqref{complex}.

Kerr BHs with scalar hair reduce to a subset of Kerr BHs in the limit of vanishing scalar field and to spinning boson stars in the limit of vanishing horizon. They can be regarded as a bound state of `bald' BHs and boson stars,  following again the paradigm of {\it ``event horizons inside classical lumps"}.
\cite{Kastor:1992qy} 
For instance they have ergo-surfaces which, in a region of parameter space, 
are a union of an ergo-sphere (like for Kerr) and an ergo-torus 
(typical of some rotating boson stars).\cite{Herdeiro:2014jaa} 
They also exhibit some distinct physical properties, 
as an independent quadrupole which can be more than one order of magnitude larger than 
that of Kerr and violation of the Kerr bound for the angular momentum.\cite{Herdeiro:2014goa,Herdeiro:2015gia} Also, unlike some spherically symmetric hairy BHs that do not admit `short hair' $i.e.$ a concentration of the scalar field very close to the horizon,\cite{Nunez:1996xv} a test field analysis suggests this model does.\cite{Hod:2014npa}

Could Kerr BHs with scalar hair form in gravitational collapse? This issue is related to their stability properties which, at the moment of writing, have not been fully clarified. But arguments have been put forward to support that some of the solutions can be very long lived.\cite{Herdeiro:2015gia} One could then imagine two scenarios for the formation of hairy BHs. The first one is that gravitational collapse would form initially a Kerr BH which would subsequently grow hair via superradiant amplification of a scalar field. An analysis using a quasi-adiabatic evolution has been used to argue that, in this case, at most around 29\% of the BH mass can be transferred to the scalar hair around the BH.\cite{Brito:2014wla} The second scenario is that gravitational collapse produces from the outset a Kerr BH with scalar hair. It was shown in the 1990s that indeed gravitational collapse of scalar fields can produce boson stars, within spherical symmetry.\cite{Seidel:1993zk,Liebling:2012fv} It would be very interesting to revisit such numerical evolutions in more general axi-symmetric setups to investigate the formation of  Kerr BHs with scalar hair.

Asymptotically flat solutions based on the condition \eqref{sync} have also been found in higher dimensions\cite{Brihaye:2014nba}, and a pioneering example was found in higher dimensional Anti-de-Sitter spacetime.\cite{Dias:2011at} Other families of analogous solutions must exist allowing the scalar field to have self-interactions (abiding Assumption 3)\cite{Herdeiro:2014pka,Herdeiro:2015gia}, some of which were recently constructed.\cite{Kleihaus:2015iea}

Let us conclude by remarking that recent work\cite{Mesic:2014yqa,Smolic:2015txa} 
discussed constraints on classes of solutions based on a violation of Assumption 2.

%%%%%%%%%%%%%%%%%%%
\section{Summary}
\label{sec_5}
%%%%%%%%%%%%%%%%%%%

In Table 1 we summarize no-scalar-hair theorems and known scalar-hairy solutions for models discussed in this review (four dimensional, asymptotically flat, with a regular geometry, on and outside the horizon). The theories are presented by their Lagrangian density, $\mathcal{S}=\int d^4x\sqrt{-g} \mathcal{L}$, and organized by a criterion of increasing complexity (which at times is arbitrary). The no-hair theorem is placed together with a reference theory, but sometimes has a more generic applicability; obviously it applies also to special cases. When there are both no-hair theorems and solutions, the latter violate some assumption of the former. Also, solutions are not repeated in more general theories. Further solutions including explicit scalar couplings to higher order curvature invariants were discussed in Section~\ref{sec_hc}.

%%%%%%%%%%%%%%%%%%%
\section{Final remarks}
\label{sec_6}
%%%%%%%%%%%%%%%%%%%
In this paper we have reviewed the status of four dimensional asymptotically  flat BHs with scalar hair in various types of scalar models coupled to gravity and without gauge (or other types of) fields. We have started in Section~\ref{sec_2} by pointing out that a test field analysis immediately shows the different behaviour of scalar fields and electromagnetic fields on a BH background. This different behaviour can be traced to the existence of a Gauss law in the latter case, but not the former. Thus, keeping scalar fields in equilibrium with a BH as a regular configuration is certainly more difficult than for gauge fields. This difficulty was formalized in a set of no-scalar-hair theorems, reviewed in Section~\ref{sec_3}. But as in any theorem, there are assumptions which, when violated allow the existence of BHs with scalar hair. These have been reviewed in Section~\ref{sec_4}.  An overview of these solutions shows that there are various 
\begin{center}
\begin{tabular}{||  c || c | c |}
\hline			
  Theory  & No-hair & Known scalar hairy BHs with  \\
  Lagrangian density $\mathcal{L}$ & theorem & regular geometry on and outside $\mathcal{H}$ \\
   & & (primary or secondary hair;  \\
   & & regularity)\\
  \hline
  \hline
  Scalar-vacuum  & Chase\cite{Chase:1970} & \\
  $\frac{1}{4}R-\frac{1}{2}\nabla_{\mu}\Phi \nabla^\mu\Phi$ &  & \\
  \hline
  Massive-scalar-vacuum  & Bekenstein\cite{Bekenstein:1972ny} & \\
  $\frac{1}{4}R-\frac{1}{2}\nabla_{\mu}\Phi \nabla^\mu\Phi-\frac{1}{2}\mu^2\Phi^2$ &  & \\
  \hline 
  Massive-complex-scalar-vacuum  &Pena-- & Herdeiro--Radu\cite{Herdeiro:2014goa,Herdeiro:2015gia}\\
  $\frac{1}{4}R
   -\nabla_\mu \Phi^* \nabla^\mu\Phi - \mu^2 \Phi^*\Phi$ &--Sudarsky\cite{Pena:1997cy}  & (primary, regular); \\  
   & & generalizations:\cite{Kleihaus:2015iea} \\
    \hline
    & Xanthopoulos-- & Bocharova--Bronnikov--Melnikov--\\
Conformal-scalar-vacuum &  --Zannias\cite{xanthopoulos:91}  & --Bekenstein (BBMB)\cite{BBM,Bekenstein:1974sf,Bekenstein:1975ts}  \\      
$\frac{1}{4}R-\frac{1}{2}\nabla_{\mu}\Phi \nabla^\mu\Phi-\frac{1}{12}R\Phi^2$ & Zannias\cite{Zannias:1994jf} & (secondary, diverges at $\mathcal{H}$); \\  
&   & generalizations:\cite{Astorino:2014mda} \\ 
\hline 
$V$-scalar-vacuum & Heusler\cite{Heusler:1992ss,HS,Heusler:1996ft} & Many, with non-positive \\  
$\frac{1}{4}R-\frac{1}{2}\nabla_{\mu}\Phi \nabla^\mu\Phi-V(\Phi)$
& Bekenstein\cite{Bekenstein:1995un} &  definite potentials:\cite{Bechmann:1995sa,Nucamendi:1995ex,Dennhardt:1996cz,Gubser:2005ih,Kleihaus:2013tba,Anabalon:2013qua,Anabalon:2012ih,Cadoni:2015gfa}.  \\ 
& Sudarsky\cite{Sudarsky:1995zg} & (typically secondary, regular) \\
\hline
$P$-scalar-vacuum & Graham-- &  \\  
$\frac{1}{4}R+P(\Phi,X)$
& --Jha\cite{Graham:2014ina} &  \\ 
\hline
 Einstein-Skyrme & & Droz--Heusler--Straumann\cite{Droz:1991cx} \\
$ \frac{1}{4}R-\frac{1}{2}\nabla_\mu \Phi^a \nabla^\mu \Phi^a$ & & (primary but topological; regular); 
\\
\ \ \ \ \ \ \ $-\kappa|\nabla_{[\mu} \Phi^a \nabla_{\nu]} \Phi^b|^2$ & & generalizations:\cite{Bizon:1992gb,Torii:1993vm} \\
 \hline
  & Hawking~\cite{Hawking:1972qk} &  \\  
Scalar-tensor theories & Saa\cite{Saa:1996aw,Saa:1996qq}  &  \\ 
$\varphi\hat{R}-\frac{\omega(\varphi)}{\varphi}\hat{\nabla}_\mu\varphi\hat{\nabla}^\mu\varphi-U(\varphi)$ & Sotiriou-- & \\ 
 & --Faraoni~\cite{Sotiriou:2011dz} & \\
\hline
 &  & Sotiriou-Zhou\cite{Sotiriou:2014pfa}  \\  
Horndeski/Galileon theories & Hui-- & (secondary; regular)  \\ 
Full $\mathcal{L}$ in eq.~\eqref{ha} & --Nicolis\cite{Hui:2012qt}   & Babichev--Charmousis\cite{Babichev:2013cya,Charmousis:2014zaa} \\ 
 & & (secondary\cite{Babichev:2013cya} or primary\cite{Charmousis:2014zaa},  \\ & & diverges at $\mathcal{H}^+$ or $\mathcal{H}^-$); \\ & & generalizations:\cite{Zhong:2015ina,Charmousis:2015aya,Babichev:2015rva} \\
\hline
\end{tabular}
\end{center}

\bigskip

\begin{center}
\noindent {\small {\bf Table 1:} Summary of no-scalar-hair theorems and asymptotically flat scalar-hairy BHs.}
\end{center}
\bigskip
\noindent possible mechanisms to construct regular asymptotically flat BHs with scalar hair. This, in particular, means that physical properties may vary substantially.

\bigskip

Nevertheless some patterns emerge. One pattern that can be observed from this overview is that scalar field theories that, when minimally coupled to gravity, allow the existence of gravitating solitons, also allow the existence of BHs with scalar hair. Three examples are: $i)$ scalarons in $V$-scalar-vacuum; $ii)$ gravitating Skyrmions in Einstein-Skyrme theory; $iii)$ boson stars in massive-complex-scalar vacuum.  In all cases the possibility of placing {\it ``event horizons inside classical lumps"} is observed, even though sometimes with subtleties. For instance, in the case of boson stars, the BHs need to have angular momentum, $cf.$ Section \ref{kbhsh}.\cite{Herdeiro:2014ima}

Another mechanism which leads generically to scalar hair on black holes
is to consider higher curvature term corrections to Einstein gravity, coupled with the scalar field.
 Note also  that,
as proven by this class of solutions  (or by the BBMB solution of conformal-scalar-vacuum)
  not all hairy solutions admit a non-trivial zero horizon limit.

Either due to the existence of gravitating solitons, or due to the existence of some particular scalar-gravity couplings, the conclusion is that despite the fact that scalar fields do not have a Gauss law and consequently are hard to keep in equilibrium with an event horizon without trivializing, BH solutions with scalar hair $exist$ nevertheless.  In that sense, the no-scalar-hair conjecture has been proved to be false. 

Still, a relevant question is: do these solutions possess new ``quantum numbers", $i.e.$ primary hair? The pattern seems to be that primary hair only occurs if the scalar field theory possesses a global symmetry and, associated with it, a conserved current and a conserved Noether charge(s). Three examples are the $O(4)$ symmetric Einstein-Skyrme theory, the shift symmetric Horndeski/Galileon theories and the $U(1)$ symmetric massive-complex-scalar vacuum. The existence of such global symmetry allows a conserved charge; but it does not guarantee that asymptotically flat, regular BH solutions with non-trivial scalar hair exist with that charge. Indeed, out of these three examples, the only case where a clear new continuous (non-topological) charge exists is the case of Kerr BHs with scalar hair of massive-complex-scalar-vacuum (and generalizations thereof).

The next question is if the \textit{dynamical} spirit of the no-hair conjecture has also been falsified. As discussed in the Introduction, the backbone idea of this conjecture is that the end point of gravitational collapse are very simple configurations without hair. Thus, dynamical stability or at the very least, the presence of only long-term instabilities, is fundamental. Some of the models described -- for instance Einstein-Skyrme -- allow physically reasonable solutions in this respect. In some other cases this not yet known, and probing stability may only be possible using numerical relativity techniques.\cite{Cardoso:2014uka} 

The final question is if any of these models is phenomenologically interesting to describe the BH candidates identified in the Cosmos. In this respect, the final word must come from matching with observations.\cite{Berti:2015itd}

%%%%%%%%%%%%%%%%%%%
\section*{Acknowledgements}
We are grateful to V. Cardoso, C. Charmousis, J.C. Degollado, M. Salgado, M. Volkov and T. Tchrakian for comments on a draft of this paper. We also thank the participants of the \textit{VIIth Black Holes Workshop}, 18-19 December 2015, Aveiro, Portugal for discussions on this topic. C.H. and E.R. thank funding from the FCT-IF programme and the grants PTDC/FIS/116625/2010,  NRHEP--295189-FP7-PEOPLE-2011-IRSES and by the CIDMA strategic funding UID/MAT/04106/2013.
%%%%%%%%%%%%%%%%%%%

%\section{References}

%\begin{thebibliography}{000} %for 3 digits
%\begin{thebibliography}{00}  %for 2 digits

\end{document}